\documentclass[11pt]{article}

\usepackage{graphics}

\title{{\bf Instabilities of dispersion-managed solitons \\
in the normal dispersion regime \vspace{2cm}}}
 
\author{Dmitry E. Pelinovsky\thanks{dmpeli@math.toronto.edu} \vspace{0.3cm}\\ 
Department of Mathematics, University of Toronto \\ Toronto, Ontario, 
Canada, M5S 3G3}

\date{\today}

\begin{document}

\baselineskip 20 pt
\maketitle
 
\begin{center}
{\bf Abstract:} 
\end{center}
\noindent
Dispersion-managed solitons are reviewed within a Gaussian variational approximation 
and an integral evolution model. In the normal regime of the dispersion map (when the 
averaged path dispersion is negative), there are two solitons of different pulse 
duration and energy at a fixed propagation constant. We show that the short soliton 
with a larger energy is linearly (exponentially) unstable. The other (long) soliton 
with a smaller energy is linearly stable but hits a resonance with excitations of 
the dispersion map. The results are compared with the results from the recent publications 
\cite{Dor1,GrMen}.

\vskip 1 cm

 
\vskip 0 cm
 

\newpage
\section{Introduction}

New ways in optimization of existing telecommunication systems based on 
dispersion management technology attracted recently wide research interest 
from soliton-based groups (see reviews \cite{Dorrev,Turrev}). The main idea 
was to combine a high local group-velocity dispersion with a low path-average 
dispersion. The former feature results in the reduction of the four-wave mixing while 
the latter one reduces the Gordon--Haus timing jitter effects. When the path-average 
dispersion is small and normal, i.e. the defocussing segment in the fiber is dominant 
over the focussing one, a new phenomenon of branching of soliton solutions 
was discovered \cite{Dor1,Dor2,GrMen,LYKM,TJ}. The soliton propagation in 
this regime is not supported by an uniform-dispersion optical fiber and seems to be 
one of the remarkable achievement of the dispersion management 
with sufficiently high local dispersion.

The stability of branching soliton solutions in the normal dispersion 
regime was a subject of intense studies which lead to contradictory 
conclusions. Grigoryan and Menyuk announced the linear and nonlinear stability of 
both the branches \cite{GrMen}, while Berntson {\em et al.} conjectured 
instability of one of the branches \cite{Dor1}. 

In this paper, we intend to shed light on the complicated issue of existence 
and stability of soliton signals in the normal regime of the dispersion map. 
We find, in the small-amplitude approximation, that there exist two branches 
of soliton solutions for different levels of energy and different pulse 
durations at a fixed propagation constant. The short pulses with larger energy 
are proved to be linearly unstable, while the other (long) pulses with smaller 
energy are neutrally stable. We show that the transition from large-energy 
unstable solitons to the stable soliton signals occurs via long-term 
transient oscillations. The two branches of soliton solutions correspond 
to a single (small-energy) branch B in Fig. 1 of \cite{GrMen}. Depending 
on a normalization condition (see Section 4), this branch may be either 
stable or unstable.

Our strategy to develop the small-amplitude approximation is based on the combination 
of two analytical approaches: the Gaussian variational approximation and the integral 
evolution model. 

The Gaussian variational approximation, being inaccurate in details, is still useful 
for a quick and rough analysis (see \cite{TS} for review and references). Also it was 
shown that the method can be extended to a rigorous Gauss-Hermite expansion of the 
basic model \cite{HGPRE}. We improve the previous results summarized in \cite{TS} 
by deriving a new dynamical system from the variational equations of a Gaussian pulse. 
The system clearly displays the linear and nonlinear instability of the short Gaussian 
pulse with larger energy. 

More rigorous analysis of the problem is based on the integral evolution model obtained 
by Gabitov and Turitsyn \cite{GT} and by Ablowitz and Biondini \cite{AB}. Although 
this model is more complicated from computational point of view (see recent papers 
\cite{TurHG,Pare}), we managed to study numerically the construction of the linear 
spectrum of dispersion-managed solitons. Our results confirm the instability and 
transition scenarios predicted within the variational model. We also deduce from 
this model that the soliton signals in the normal regime of the dispersion map are 
in resonance with the wave continuum of linear excitations of the map. The resonance 
implies isually the generation of wave packets from stable pulses oscillating in time. 
The latter effects are beyond the accuracy of the analytical model and are left 
for further studies. 

\section{Gaussian approximation: New dynamical model}

We study the NLS model in the dimensionless form \cite{LYKM},
\begin{equation}
\label{NLS}
i u_z + \frac{1}{2} D(z) u_{tt} + \epsilon \left( \frac{1}{2} 
D_0 u_{tt} + |u|^2 u \right) = 0,
\end{equation}
where $u(z,t)$ is the envelope of an optical pulse in the retarded reference frame 
of the fiber. The small parameter $\epsilon$ measures the length of the dispersion's 
map and the inverse variance of the local dispersion. After normalization, $D_0$ and 
$D(z)$ are assumed to be of order of ${\rm O}(1)$, and
\begin{equation}
\label{D}
\langle D \rangle = \int_0^1 D(z) dz = 0, \;\;\;\;\;
D(z+1) = D(z).
\end{equation}
Further physical motivations for derivation and verification of the model 
(\ref{NLS}) can be found in \cite{Dorrev,Turrev}. Soliton-like optical 
pulses are solutions of the model in the form,
\begin{equation}
\label{solitonNLS}
u(z,t) = \psi(z,t) e^{i \epsilon \mu z},
\end{equation}
where $\mu$ is the propagation constant and $\psi(z,t)$ is a soliton 
profile satisfying the boundary conditions, 
\begin{equation}
\label{boundary1}
\psi(z+1,t) = \psi(z,t)
\end{equation}
and 
\begin{equation} 
\label{boundary2}
\lim_{t \to \infty} \psi(z,t) = 0.
\end{equation}

One of the conventional approximation for soliton solutions of NLS-type 
equations is based on averaging the Gaussian anzatz in the Lagrangian 
density and further varying the Lagrangian density with respect to 
parameters of the Gaussian pulse (see \cite{TS} for review).  The Gaussian 
approximation is the first term of the Gauss-Hermite expansions when  
solving the NLS equation (\ref{NLS}) in the limit $\epsilon \to 0$ 
\cite{HGPRE}. We apply the Gaussian anzatz in the form,
\begin{equation}
\label{Gaussian}
u(z,t) = c(z) \exp\left(- \frac{(\alpha(z) - 2 i \beta(z))}{\alpha(z)^2 + 
4 \beta(z)^2} t^2 + i \phi(z)\right).
\end{equation}
Here the four parameters of the Gaussian pulse are: $c(z)$ - the amplitude, $\phi(z)$ 
- the gauge parameter, $\alpha(z)$ - the pulse duration, and $\beta(z)$ - the chirp. 
It was found that the four equations for variations of the Lagrangian density can be 
decoupled into a system for $\alpha(z)$ and $\beta(z)$ of the form, 
\begin{eqnarray}
\label{alpha}
\frac{d \alpha}{d z} & = & \frac{4 \epsilon E \alpha^{5/2} \beta}{
(\alpha^2 + 4 \beta^2)^{3/2}}, \\
\label{beta}
\frac{d \beta}{d z} & = & D(z) + \epsilon \left( D_0 - 
\frac{E \alpha^{3/2} ( \alpha^2 - 4 \beta^2)}{
2 (\alpha^2 + 4 \beta^2)^{3/2}} \right).
\end{eqnarray}
The phase factor $\phi(z)$ is expressed in terms of $\alpha(z)$ and $\beta(z)$,
\begin{equation}
\frac{d}{d z} \left( \phi + \frac{1}{2} \arctan \frac{2 \beta}{\alpha} \right)
= \frac{\epsilon E \alpha^{1/2} ( 3 \alpha^2 + 20 \beta^2)}{
4 (\alpha^2 + 4 \beta^2)^{3/2}},
\end{equation}
while the amplitude $c(z)$ is given in terms of the input energy constant $E$ as
\begin{equation}
\label{energy}
E = \frac{\sqrt{\alpha^2 + 4 \beta^2}}{\sqrt{2 \alpha}} c^2 = 
\frac{1}{\sqrt{\pi}} \int_0^1 dz \int_{-\infty}^{\infty} dt |u|^2(z,t) > 0.
\end{equation}
The stationary pulse (\ref{solitonNLS}) - (\ref{boundary2}) corresponds to the 
periodic solutions of the system (\ref{alpha}) and (\ref{beta}) in the form, 
\begin{equation}
\label{periodic}
\alpha(z+1) = \alpha(z), \;\;\;\;
\beta(z+1) = \beta(z), \;\;\;\;
\phi(z+1) = \phi(z) + \epsilon \mu.
\end{equation}
For simplicity, we study the periodic solutions in the limit $\epsilon \to 0$ 
by using a two-step dispersion map with zero average,
\begin{equation}
\label{two_step_map}
D(z) = \left\{ \begin{array}{c} D_1, \;\;\;\;0<z<L \\
D_2, \;\;\;\;L<z<1 \end{array} \right. ,
\end{equation}
where 
$$
m = D_1 L = - D_2 (1 - L) > 0.
$$
The asymptotic solution in the limit $\epsilon \to 0$ can be sought in the 
regular form,
$$
\alpha(z) = \alpha_s + {\rm O}(\epsilon), \;\;\;\;
\beta(z) = \int_0^z D(z') dz' + \beta_s + {\rm O}(\epsilon),
$$
where $\alpha_s$, $\beta_s$ are constant. The periodic solutions 
appear when $\beta_s = - m/2$ and $\alpha_s$ is a root of the equation,
\begin{equation}
\label{root1}
D_0 = E \alpha_s^{3/2} \left[ \frac{1}{(m^2 + \alpha_s^2)^{1/2}} - 
\frac{1}{2 m} \log \left( \frac{m+(m^2 + \alpha_s^2)^{1/2}}{\alpha_s} 
\right) \right].
\end{equation}
In addition, the propagation constant $\mu$ can be obtained as a function 
of $E$ and $\alpha_s$ according to the equation,
\begin{equation}
\label{root2}
\mu = \frac{1}{4} E \alpha_s^{1/2} \left[ \frac{-2}{(m^2 + \alpha_s^2)^{1/2}} 
+ \frac{5}{m} \log \left( \frac{m+(m^2 + \alpha_s^2)^{1/2}}{\alpha_s} 
\right) \right].
\end{equation}
These equations have been already derived in the literature, see \cite{LYKM,TJ} 
for (\ref{root1}) and \cite{HGPRE} for (\ref{root2}). However, the relations 
(\ref{root1}) and (\ref{root2}) were viewed typically under the normalization condition, 
\begin{equation}
\label{normalization}
m = 1, \;\;\;\; E = \frac{1}{\sqrt{2 S}}, \;\;\;\; \alpha_s = \frac{1}{S},
\end{equation}
where $S$ is called the map strength. In this normalization, the expression (\ref{root1}) 
gives a small-amplitude limit of the results of \cite{Dor1,GrMen}, i.e. the 
slope $E/D_0$ is a function of $S$. The existence of solitons 
was identified both for $D_0 > 0$ (when $S < S_{thr}$) and for $D_0 < 0$ (when 
$S < S_{thr}$), where $S_{thr} \approx 3.32$. 

In this paper, we develop a different frame to view the soliton solutions 
(\ref{root1}) and (\ref{root2}). Guided by the stability analysis of solitons 
in generalized NLS equations \cite{DimaPD}, we fix the parameters of the model 
($D_0,m$) and construct periodic solutions as a one-parameter family in terms 
of the propagation constant $\mu$. As a result, the parameters $\alpha$ and $E$ 
can be found from (\ref{root1})-(\ref{root2}) as $\alpha_s = \alpha_s(\mu)$ and 
$E = E(\mu)$. These functions are shown in Fig. 1(a,b) for $D_0 = 0.02$ and $m = 2$ 
and in Fig. 2(a,b) for $D_0 = -0.02$ and $m = 2$. Obviously, the branching occurs 
at $D_0 < 0$ (i.e. at the normal regime of the dispersion map), when the dispersion 
map is defocussing on average. The two solutions coexist for 
$\mu > \mu_{thr}(D_0,m)$ and $E > E_{thr}(D_0,m)$. Both the branches I and II 
correspond to a single branch B in the small-amplitude approximation under the 
normalization condition (\ref{normalization}) \cite{GrMen}.

In order to describe non-stationary dynamics of the Gaussian pulse near the periodic 
solutions, we derive a dynamical model from Eqs. (\ref{alpha}) and (\ref{beta}) by 
setting, 
$$
\alpha(z) = \alpha_0(\zeta) + \epsilon \alpha_1(z,\zeta) + {\rm O}(\epsilon^2)
$$
and 
$$
\beta(z) = \int_0^z D(z') dz' + \beta_0(\zeta) + \epsilon \beta_1(z,\zeta) 
+ {\rm O}(\epsilon^2).
$$
Here $\zeta = \epsilon z$ is the distance to measure the evolution of a Gaussian 
pulse over many map's periods. The coupled system (\ref{alpha}) and 
(\ref{beta}) can be averaged over the map's period subject to the periodic 
conditions: $\alpha_1(z+1,\zeta) = \alpha_1(z,\zeta)$ and $\beta_1(z+1,\zeta) 
= \beta_1(z,\zeta)$. Then, the non-stationary system reduces to the dynamical 
model for $\alpha_0(\zeta)$ and $\beta(\zeta)$, 
\begin{eqnarray}
\label{alpha0}
\frac{d \alpha_0}{d \zeta} = F_{\alpha}(\alpha_0,\beta_0) & \equiv & 
\frac{E \alpha_0^{5/2}}{m} \left[ \frac{1}{(\alpha_0^2 + 4 \beta_0^2)^{1/2}} - 
\frac{1}{(\alpha_0^2 + 4 (\beta_0+m)^2)^{1/2}} \right], \\
\nonumber 
\frac{d \beta_0}{d \zeta} = F_{\beta}(\alpha_0,\beta_0) & \equiv & D_0 - 
\frac{E \alpha_0^{3/2}}{4 m} \left[ \frac{4(m+\beta_0)}{(\alpha_0^2 
+ 4 (m + \beta_0)^2)^{1/2}} - \frac{4 \beta_0}{(\alpha_0^2 + 
4 \beta_0^2)^{1/2}} \right. \\
\label{beta0}
& + & \left.
\log \left( \frac{2 \beta_0 + (\alpha_0 + 
4 \beta_0^2)^{1/2}}{2 (m+\beta_0) + (\alpha_0^2 + 4 (m + 
\beta_0)^2)^{1/2}} \right) \right].
\end{eqnarray}
This system has of course the same stationary solutions $\alpha_0 = \alpha_s$ 
and $\beta_0 = \beta_s = - m/2$ as those given in (\ref{root1}). The stationary 
solutions appear as equilibrium states in the dynamical system, whose 
stability can be found by linearizing, 
$$
\alpha_0(\zeta) = \alpha_s + \Delta \alpha e^{i \lambda \zeta}, 
$$
$$
\beta_0(\zeta) = \beta_s + \Delta \beta e^{i \lambda \zeta},
$$
where the eigenvalue $\lambda$ is
\begin{equation}
\label{lambda}
\lambda^2(\mu) = - \frac{\partial F_{\alpha}}{\partial \beta}(\alpha_s) 
\frac{\partial F_{\beta}}{\partial \alpha}(\alpha_s) = 
\frac{2 E \alpha_s^{3/2}}{(m^2 + \alpha_s^2)^{3/2}} \left[ 
3 D_0 + \frac{E \alpha_s^{3/2} (m^2 - \alpha_s^2)}{(m^2 + \alpha_s^2)^{3/2}} 
\right].
\end{equation}
We plot $\lambda^2(\mu)$ in Fig.3 to confirm that $\lambda^2 > 0$ for branch I 
of the periodic solutions and $\lambda^2 < 0$ for branch II (cf. Fig. 2). Thus, 
the linear analysis predicts the instability of the short Gaussian pulses with 
larger energy at a fixed propagation constant $\mu$ (branch II). In the limit 
$\mu \to \mu_{thr}(D_0,m)$, the instability disappears, i.e.  
$$
\lim_{\mu \to \mu_{thr}} \lambda^2(\mu) = 0.
$$
This property follows from Eq. (\ref{beta0}) in the limit $\mu \to \mu_{thr}(D_0,m)$, 
when 
\begin{equation}
\label{lambda1}
\frac{\partial F_{\beta}}{\partial E}(\alpha_s,\beta_s) \frac{d E}{d \mu} + 
\frac{\partial F_{\beta}}{\partial \alpha}(\alpha_s,\beta_s) \frac{d \alpha}{d \mu} = 0.
\end{equation}
Connecting Eqs. (\ref{lambda}) and (\ref{lambda1}), we find the following 
asymptotic approximation,
\begin{equation}
\label{lambda2}
\lambda^2 \to \left( \frac{\partial F_{\beta}/\partial E \cdot 
\partial F_{\alpha}/\partial \beta}{d\alpha/d \mu} \right) \frac{d E}{d \mu}.
\end{equation}
Taking into account that $d E / d \mu$, $\partial F_{\beta}/\partial E$, and $\partial 
F_{\alpha}/\partial \beta$ are all positive for $D_0 < 0$, and $\alpha \sim 
(\mu - \mu_{thr})^{1/2}$ (see Fig. 2(a,b)), the asymptotic approximation 
(\ref{lambda2}) produces the result, $\lambda \sim (\mu - \mu_{thr})^{1/4}$. 

\begin{figure}
\begin{minipage}{10cm}
\begin{center} \hspace*{-0.5cm} (a) \end{center}
\vspace{-2cm}
\rotatebox{0}{\resizebox{8cm}{10cm}{\includegraphics[0in,0.5in]
 [8in,10.5in]{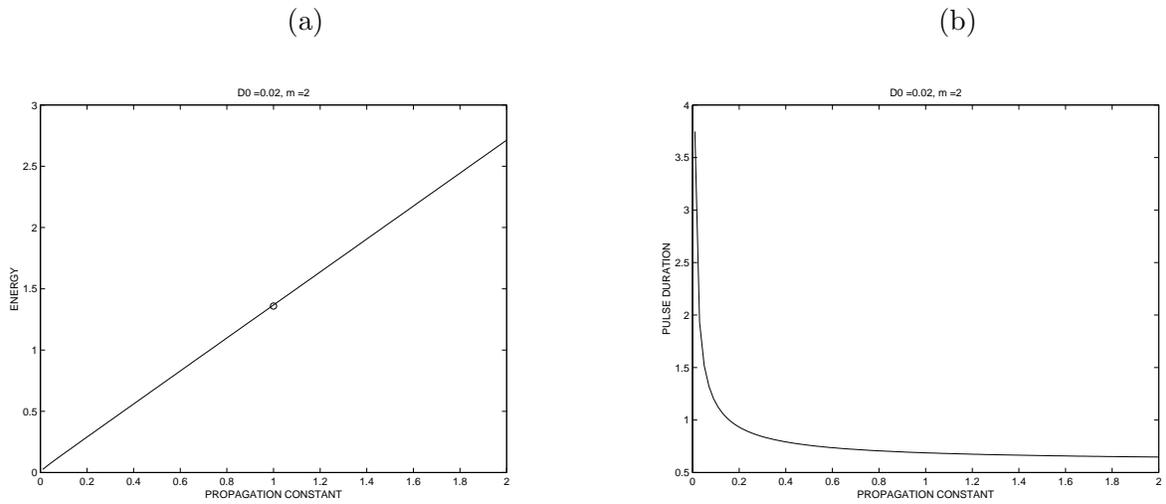}}}
\end{minipage}
\hspace{-1.5cm}
\begin{minipage}{10cm}
\begin{center} \hspace*{-0.5cm} (b) \end{center}
\vspace{-2cm}
\rotatebox{0}{\resizebox{8cm}{10cm}{\includegraphics[0in,0.5in]
 [8in,10.5in]{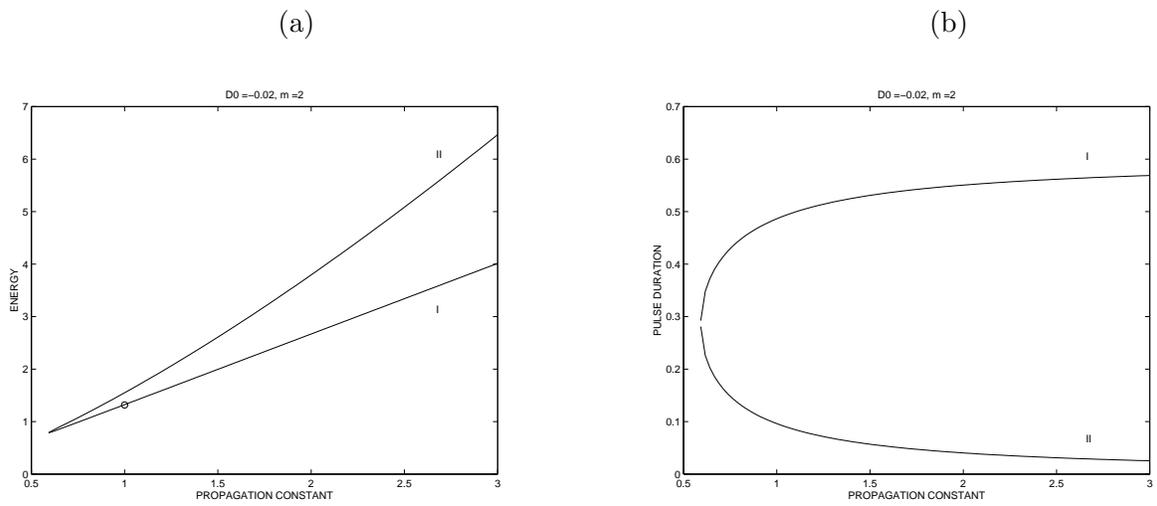}}}
\end{minipage}
\vspace{-1.5cm}
\caption{Energy $E$ (a) and pulse duration $\alpha_s$ (b) for 
the Gaussian pulse (\ref{Gaussian}) at $D_0 = 0.02$, $m = 2$.}
\end{figure}

\begin{figure}
\begin{minipage}{10cm}
\begin{center} \hspace*{-0.5cm} (a) \end{center}
\vspace{-2cm}
\rotatebox{0}{\resizebox{8cm}{10cm}{\includegraphics[0in,0.5in]
 [8in,10.5in]{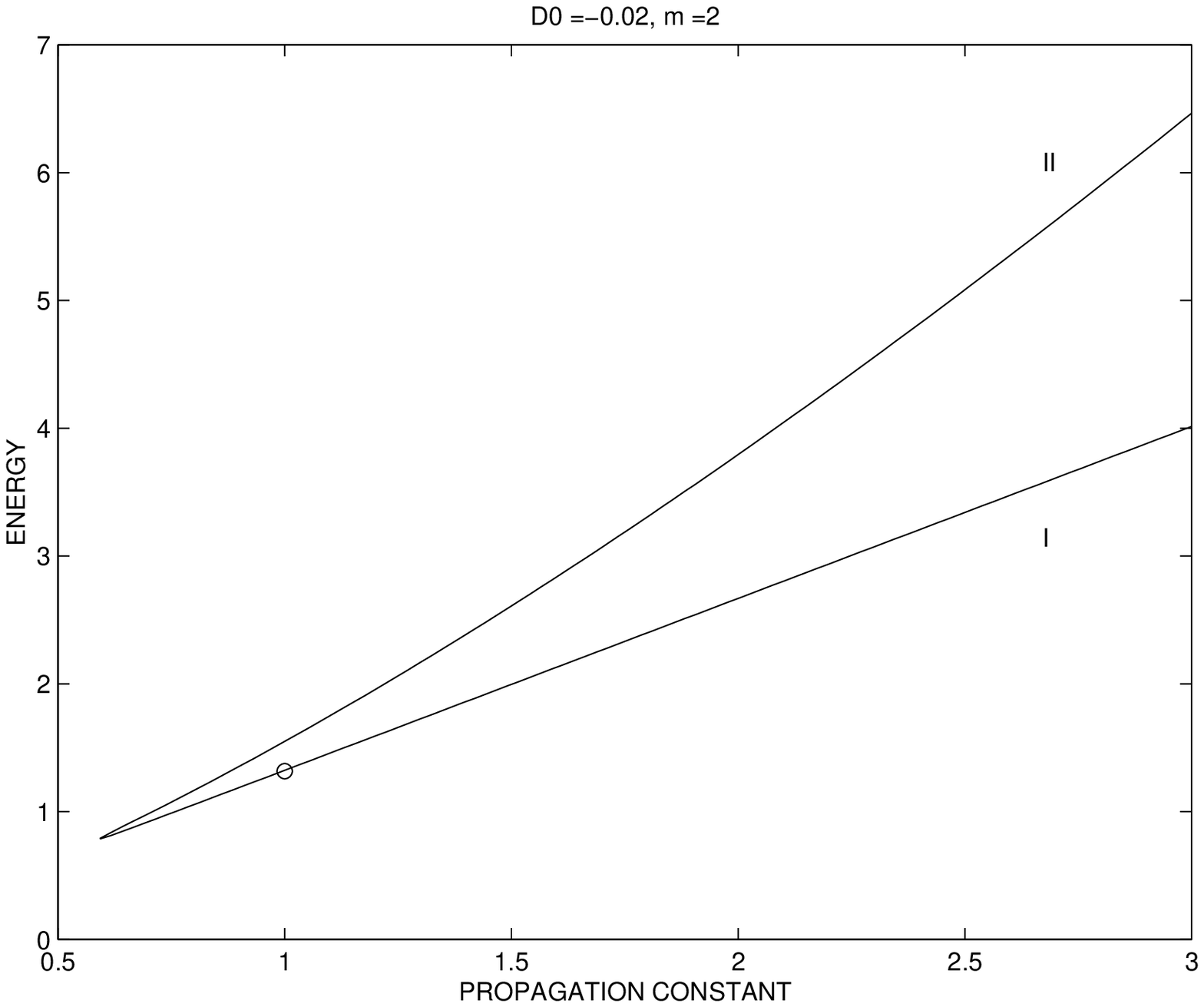}}}
\end{minipage}
\hspace{-1.5cm}
\begin{minipage}{10cm}
\begin{center} \hspace*{-0.5cm} (b) \end{center}
\vspace{-2cm}
\rotatebox{0}{\resizebox{8cm}{10cm}{\includegraphics[0in,0.5in]
 [8in,10.5in]{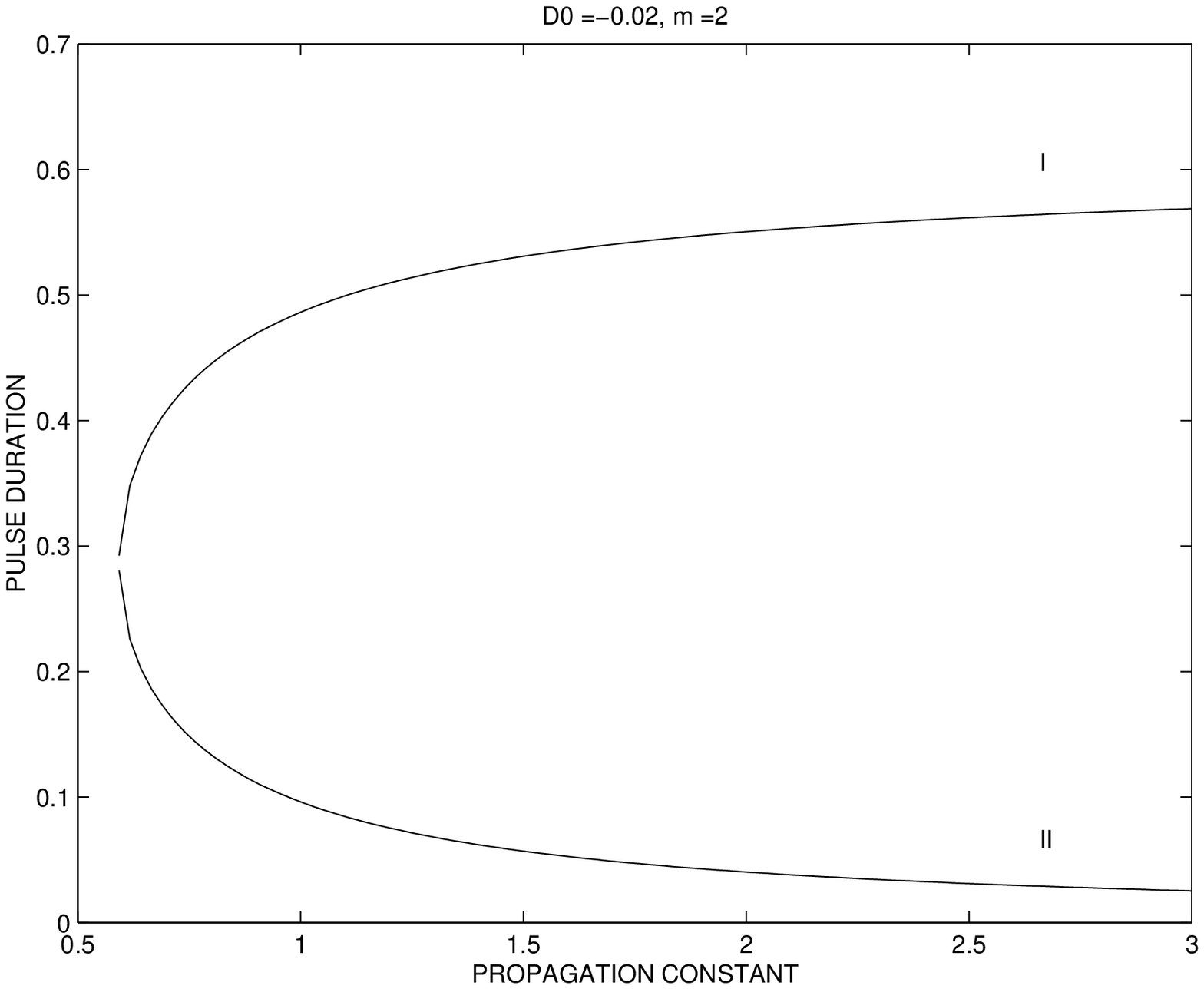}}}
\end{minipage}
\vspace{-1.5cm}
\caption{Energy $E$ (a) and pulse duration $\alpha_s$ (b) for 
the Gaussian pulse (\ref{Gaussian}) at $D_0 = -0.02$, $m = 2$.}
\end{figure}

The nonlinear dynamics of the system (\ref{alpha0}) and (\ref{beta0}) 
is shown in Fig. 4 for $D_0 = -0.02$ and $m = 2$. At a fixed value of 
the energy $E$, there are two stationary Gaussian pulses of different durations: 
a short pulse is a saddle point, while a long one is a center. Inside the 
separatrix loop, there are oscillations of the pulse trapped by the center 
point. Outside the separatrix, the Gaussian pulse transfers to chirped 
linear waves. 

\begin{figure}
\begin{minipage}{10cm}
\vspace{-2cm}
\rotatebox{0}{\resizebox{10cm}{10cm}{\includegraphics[0in,0.5in][8in,11in]{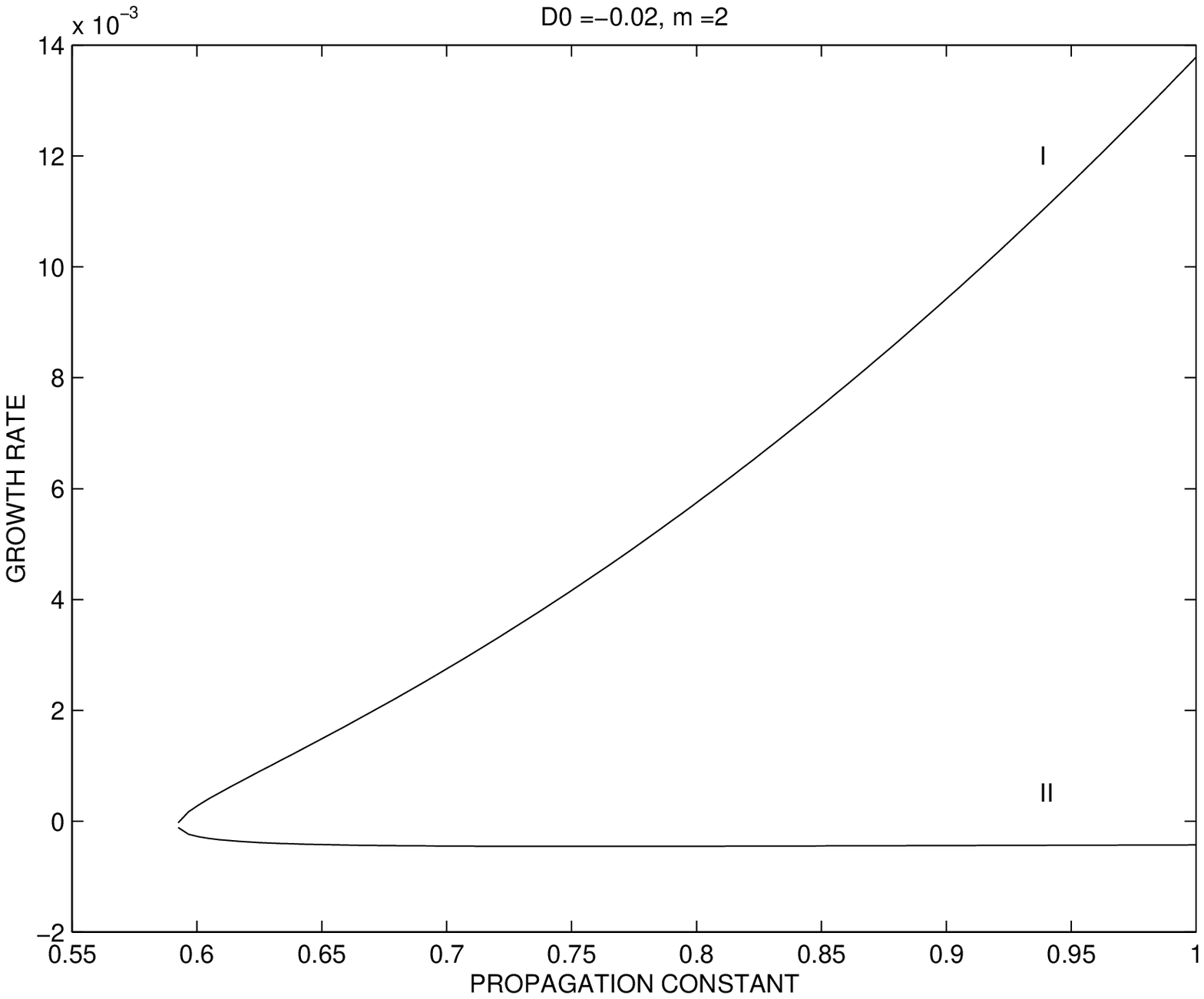}}}
\end{minipage}
\vspace{-1.5cm}
\caption{Growth rate $\lambda^2$ for the Gaussian pulse (\ref{Gaussian}) 
at $D_0 = -0.02$, $m = 2$. The branch II with $\lambda^2 < 0$ corresponds 
to the unstable Gaussian pulse.}
\begin{minipage}{10cm}
\vspace{-2cm}
\rotatebox{0}{\resizebox{10cm}{10cm}{\includegraphics[0in,0.5in][8in,11in]{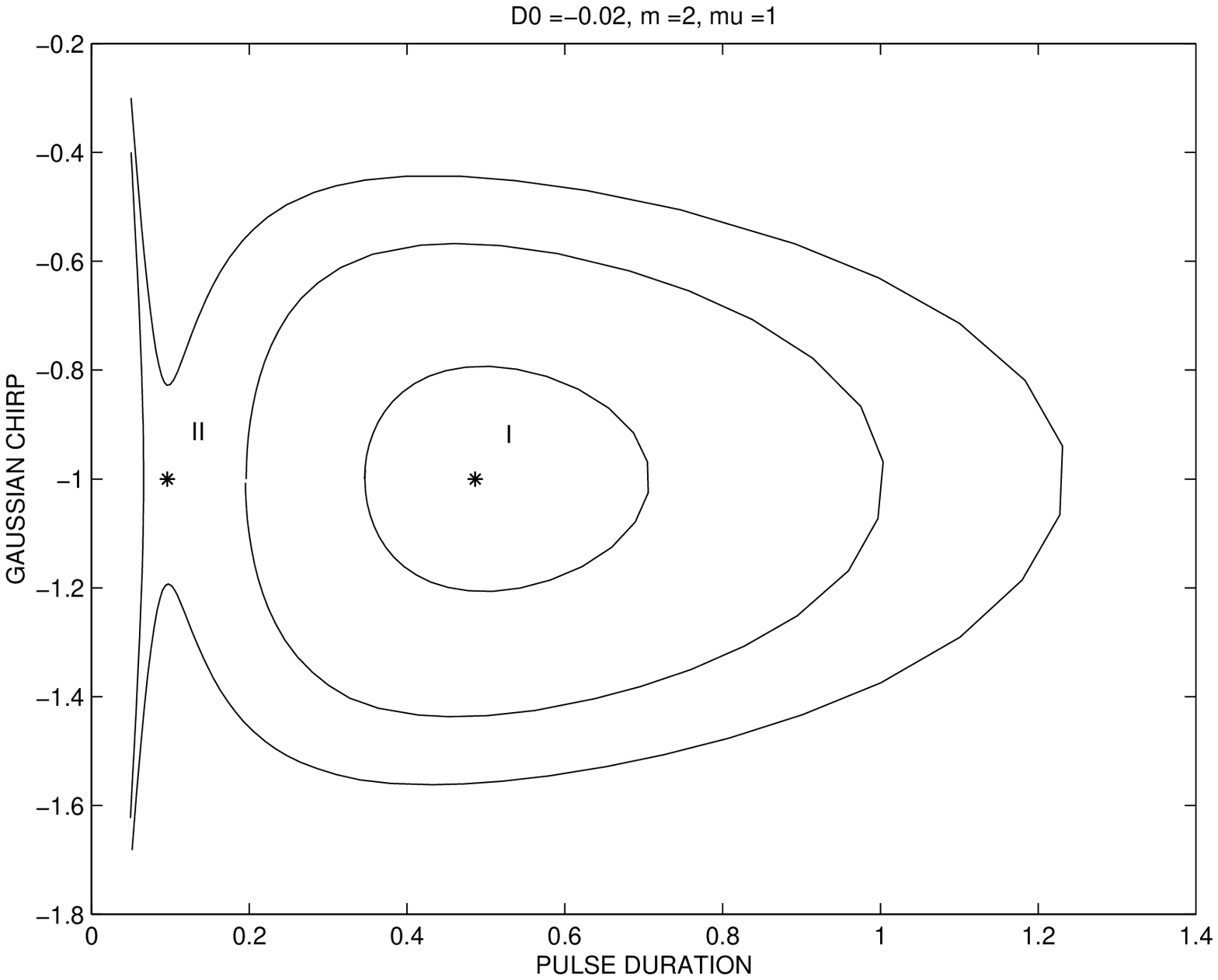}}}
\end{minipage}
\vspace{-1.5cm}
\caption{Phase plane ($\beta$, $\alpha$) for the nonlinear dynamics of a Gaussian 
pulse (\ref{Gaussian}) at $D_0 = -0.02$, $m = 2$ and $\mu = 1$.}
\end{figure}

We notice that the transition scenario resembles the nonlinear dynamics of unstable 
solitons in generalized NLS equations \cite{DimaPD}. The only difference is that 
the unstable branch in generalized NLS equations is located for those values of 
soliton propagation constant $\mu$, where $d E / d\mu < 0$. Although this conventional 
stability criterion failed for the dispersion-managed solitons (see Fig. 2(a)), 
the instability 
development shows up to be alike (cf. Fig. 4 here and Fig. 2(b) in \cite{DimaPD}).

\section{Integral evolution model: Numerical analysis}

The Gaussian approximation of the optical pulse in the NLS model 
(\ref{NLS}) can be improved by summating all higher-order Gauss-Hermite 
solutions of the linear equation, $i u_z + 0.5 D(z) u_{tt} = 0$ as 
shown in \cite{HGPRE}. However, this analysis results in a complicated 
infinite-dimensional system of algebraic equations for parameters and 
coefficients of the Gauss-Hermite expansion. Instead, we adopt a direct 
asymptotic method to deduce an integral evolution model valid in the 
limit $\epsilon \to 0$. This method is based on Fourier expansion of 
solutions of the linear equation above \cite{GT,AB} as well as 
on the asymptotic expansion, 
$$
u(z,t) = u_0(z,t) + \epsilon u_1(z,t) + {\rm O}(\epsilon^2),
$$
where $u_0(z,t)$ is given in the Fourier form as 
\begin{equation}
\label{Fourier}
u_0 = \frac{1}{2\pi} \int_{-\infty}^{\infty} d\omega W(\omega,\zeta) 
\exp\left(- \frac{i}{2} \omega^2 \left(\int_0^z D(z')dz' \right) 
+ i \omega t \right).
\end{equation}
Here $W(\omega,\zeta)$ is a complex Fourier coefficient and $\zeta = \epsilon z$ 
is the distance to measure the pulse evolution over many map's periods. By supplying 
the periodic condition $u_1(z+1,t) = u_1(z,t)$ in the Fourier form, the NLS equation 
(\ref{NLS}) can be reduced to the integral evolution model, 
\begin{equation}
\label{integral_model}
i W_{\zeta} - \frac{1}{2} D_0 \omega^2 W + \int\!\int_{-\infty}^{\infty} 
d\omega_1 d\omega_2 r(\omega_1 \omega_2) W(\omega+\omega_1) W(\omega+\omega_2) 
\bar{W}(\omega+\omega_1+\omega_2) = 0,
\end{equation}
where 
$$
r(\omega_1 \omega_2) = \frac{1}{4 \pi^2} \int_0^1 dz \exp\left( i \omega_1 \omega_2 
\int_0^z D(z')dz' \right).
$$
For the two-step dispersion map (\ref{two_step_map}), the integral 
kernel $r(x)$ can be computed explicitly as 
\begin{equation}
\label{kernel}
r(x) = \frac{1}{4 \pi^2} \frac{\sin\left(\frac{m x}{2}\right)}{\frac{mx}{2}}.
\end{equation}
It is obvious that the dynamical system (\ref{alpha0}) and (\ref{beta0}) 
studied in the previous section can be deduced from (\ref{integral_model}) 
within the same Gaussian approximation. This correspondence implies that the 
qualitative results on instability of short Gaussian pulses for $D_0<0$ 
can be reconfirmed within a more systematic theory.

In this section we present numerical results consisting of three subsections. 
In the first subsection, we construct a numerical solution of the stationary 
problem identifying optical solitons in the normal regime, when $D_0 < 0$. 
In the second part, we analyze the linearized problem and locate numerically 
the linear spectrum in the problem, indicating possible instability of 
optical solitons. Then, we simulate numerically the non-stationary problem described 
by (\ref{integral_model}) and discuss the transformation routes of 
the unstable dispersion-managed solitons. 

\subsection{Stationary solutions}

The periodic-type localized solutions of the NLS equation in the form 
(\ref{solitonNLS}) are equivalent to stationary solutions of 
(\ref{integral_model}) in the form,
\begin{equation}
\label{soliton}
W(\omega,\zeta) = \Phi(\omega) e^{i \mu \zeta},
\end{equation}
where $\Phi(\omega)$ is the real Fourier coefficient which defines 
$\psi(z,t)$ according to (\ref{Fourier}). This function satisfies 
a nonlinear integral boundary-value problem,
\begin{equation}
\label{nonlinear}
\left( \mu + \frac{1}{2} D_0 \omega^2 \right) \Phi(\omega) = R(\omega) \equiv 
\int\!\int_{-\infty}^{\infty} d\omega_1 d\omega_2 r(\omega_1 \omega_2) 
\Phi(\omega+\omega_1) \Phi(\omega+\omega_2) \Phi(\omega+\omega_1+\omega_2), 
\end{equation}
where $\Phi(-\omega) = \Phi(\omega)$ (the symmetry condition) and 
$\lim_{\omega \to \infty} \Phi(\omega) = 0$ (the boundary condition). 

For numerical analysis, we intend to use the Petviashvili's iteration 
scheme \cite{Pet}:
$$
\Phi^{(n)}(\omega) \to \Phi^{(n+1)}(\omega)
$$
for $n=0,1,2,...$. Within this scheme, the right-hand-side $R(\omega)$ can be 
approximated at the $n$-th approximation by $\Phi^{(n)}(\omega)$ provided a 
certain stabilizing factor is introduced for convergence (see (\ref{map}) and 
(\ref{factor}) below). However, the numerical scheme breaks down for $D_0 < 0$ 
due to resonances at $\omega = \pm \omega_{res}$, where 
\begin{equation}
\label{resonance}
\omega_{res} = \sqrt{\frac{2 \mu}{|D_0|}}.
\end{equation}
Indeed for $\omega = \pm \omega_{res}$, the left-hand-side of (\ref{nonlinear}) 
vanishes. [Here we notice that $\mu > 0$ for the Gaussian pulse solutions 
(\ref{Gaussian}) of the NLS model (\ref{NLS}).] In order to avoid resonances 
in the numerical scheme, we add and subtract a dummy positive dispersion term 
$0.5 |D_0| \omega^2 \Phi(\omega)$ to the left-hand-side of (\ref{nonlinear}). 
As a result, the scheme converts to the following map,
\begin{equation}
\label{map}
\Phi^{(n)}(\omega) \to \Phi^{(n+1)}(\omega) = S_n^{3/2} \left( \frac{
R^{(n)}(\omega) + \frac{1}{2} (|D_0| - D_0) \omega^2 \Phi^{(n)}(\omega)}{
\mu + \frac{1}{2} |D_0| \omega^2 } \right), \;\;\;\; D_0 < 0,
\end{equation}
where $S_n$ is the Petviashvili's stabilizing factor given by
\begin{equation}
\label{factor}
S_n = \frac{\int_{-\infty}^{\infty} d \omega (\mu + \frac{1}{2} |D_0| \omega^2) 
\Phi^{(n)}(\omega)}{\int_{-\infty}^{\infty} d\omega \Phi^{(n)}(\omega) 
\left( R^{(n)}(\omega) + \frac{1}{2} (|D_0| - D_0) \omega^2 \Phi^{(n)}(\omega) \right)}.
\end{equation}
The factor $S_n$ is unity at the stationary solution and serves therefore 
as an indicator for termination of the iterating procedure. We stop 
iterations when $|S_n - 1| < 10^{-5}$.  

To use the map (\ref{map}), we apply the Simpson's integration method, reducing 
complexity due to the symmetry: $\Phi(-\omega) = \Phi(\omega)$. As a starting 
solution, the profile $\Phi(\omega)$ can be approximated by the Gaussian pulse 
with parameters corresponding to the periodic solution (\ref{root1}) and (\ref{root2}),
\begin{equation}
\label{initial}
\Phi^{(0)}(\omega) = \sqrt{\pi E \sqrt{2 \alpha_s}} \exp\left(
- \frac{1}{4} \alpha_s \omega^2\right).
\end{equation}

\begin{tabular}{|l|l|l|l|}
\hline
Number of iterations & $S_n$: $D_0 = 0.02$ & $S_n$: $D_0 = -0.02$ (I) & $S_n$: 
$D_0 = -0.02$ (II) \\ \hline
1 & 0.9897 & 0.9957 & 0.9920 \\ \hline
2 & 0.9971 & 0.9981 & 0.9917 \\ \hline
3 & 0.9994 & 0.9988 & 0.9921 \\ \hline
4 & 0.9998 & 0.9991 & 0.9942 \\ \hline
5 & 0.9999 & 0.9993 & 0.9989 \\ \hline
6 & & 0.9994 & 1.0069 \\ \hline
7 & & 0.9995 & 1.0167 \\ \hline
8 & & 0.9996 & 1.0173 \\ \hline
9 & & 0.9997 & 0.9642 \\ \hline
10 & & 0.9997 & 0.8133 \\ \hline
11 & & 0.9998 & 0.7025 \\ \hline
12 & & 0.9998 & 0.6833 \\ \hline
13 & & 0.9999 & 0.6773 \\ \hline
14 & & 0.9999 & 0.6695 \\ \hline
15 & & 0.9999 & 0.6601 \\ \hline
16 & & & 0.6499 \\ \hline
17 & & & 0.6394 \\ \hline
18 & & & 0.6292 \\ \hline
19 & & & 0.6192 \\ \hline
\end{tabular}
\begin{center}
Table I. Iterations of the stabilizing factor $S_n$ for $D_0 = 0.02$ and 
$D_0 = -0.02$ (branches I and II).
\end{center}

\vspace{1cm}

Table 1 shows iterations for the stabilizing factor $S_n$ in the three 
different cases: (i) $D_0 = 0.02$, (ii) $D_0 = -0.02$ (branch I), and 
(iii) $D_0 = -0.02$ (branch II). For all the cases, the other parameters 
are $\mu = 1$ and $m = 2$. In the first case, the convergence is 
monotonic and the profile for stationary soliton $\Phi(\omega)$ is shown 
in Fig. 5(a). The numerical value for energy of the stationary soliton 
is shown in Fig. 1(a) by a bullet. In the second case, the iterations 
converge slowly to the stationary soliton shown in Fig. 5(b). Sometimes, 
the convergence is accompanied by a single oscillation of $S_n$ near unity.
The numerical value for the energy is shown in Fig. 2(a) by a bullet at branch I. 
In the last case, however, the iterations oscillate and finally diverge. 
Inspection of the profile $\Phi^{(n)}(\omega)$ at a final iteration shows that 
the iterations change the initial pulse drastically leading to its 
disappearance. Two conjectures follow from this fact. Either the 
shorter soliton with larger energy at branch II does not exist as a 
stationary solution of (\ref{nonlinear}) or it is unstable within the 
iterational scheme (\ref{map}). Since the short Gaussian pulse does exist 
(see Fig. 2), we incline to work along the second conjecture. The iterational 
scheme (\ref{map}) is not relevant for the time-evolution problem and rigorous 
analysis of linearized problem is needed to confirm predictions of the instability 
of the short stationary pulse. 

\begin{figure}
\begin{minipage}{10cm}
\begin{center} \hspace*{-0.5cm} (a) \end{center}
\vspace{-2cm}
\rotatebox{0}{\resizebox{8cm}{10cm}{\includegraphics[0in,0.5in]
 [8in,10.5in]{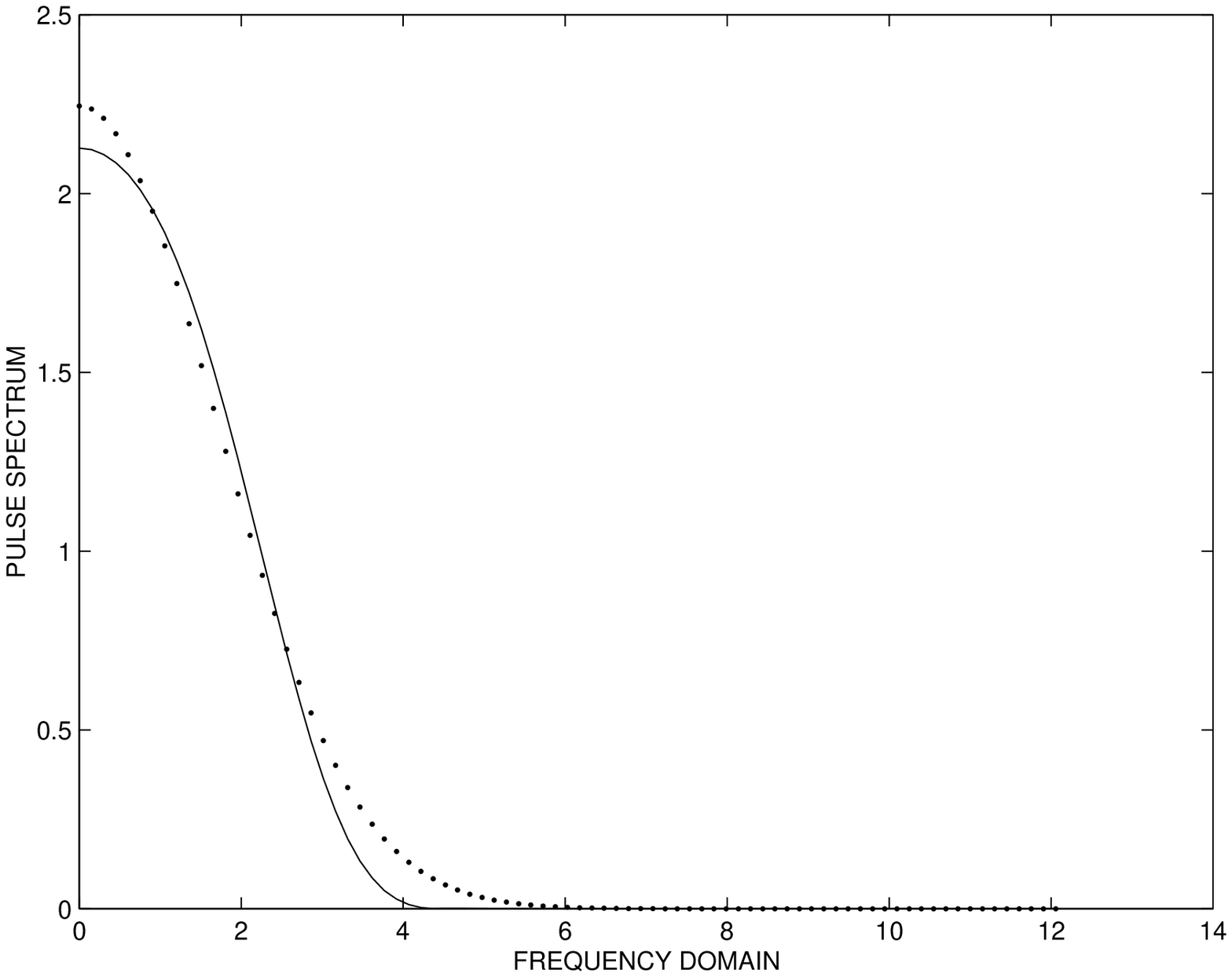}}}
\end{minipage}
\hspace{-1.5cm}
\begin{minipage}{10cm}
\begin{center} \hspace*{-0.5cm} (b) \end{center}
\vspace{-2cm}
\rotatebox{0}{\resizebox{8cm}{10cm}{\includegraphics[0in,0.5in]
 [8in,10.5in]{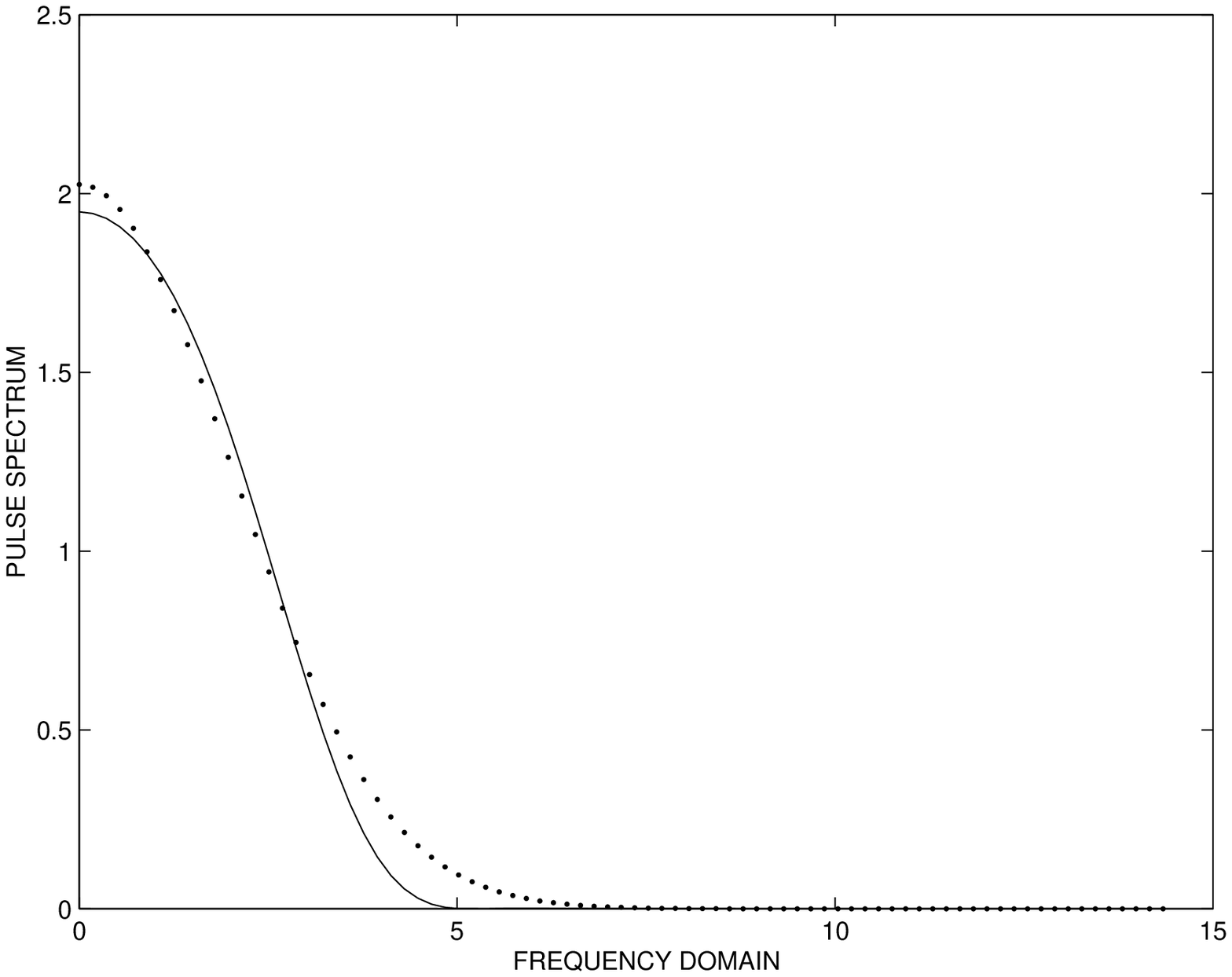}}}
\end{minipage}
\vspace{-1.5cm}
\caption{The profile of the stationary pulse $\Phi(\omega)$ in the boundary-value 
problem (\ref{nonlinear}) at $D_0 = 0.02$ (a) and $D_0 = -0.02$ (branch I) (b). 
The other parameters are: $m = 2$ and $\mu = 1$. The dotted line displays the 
Gaussian pulse (\ref{initial}) for the same parameter values.}
\end{figure}

\subsection{Linear spectrum}

There are several forms of the linear problem associated to the 
NLS-type equations. We will use the matrix form which was studied in 
our previous paper \cite{prev} subject to certain simplifications. 
The matrix form appears upon perturbations of the stationary solutions 
of (\ref{integral_model}) as 
\begin{equation}
\label{linearization}
W(\omega,\zeta) = e^{i \mu \zeta} \left[ \Phi(\omega) + w_1(\omega) 
e^{i \lambda \zeta} + \bar{w}_2(\omega) e^{-i \lambda \zeta} \right].
\end{equation}
The vector ${\bf w}(\omega) = (w_1,w_2)^T$ can be shown to satisfy the 
matrix linear problem, 
\begin{equation}
\label{linear_problem}
\lambda {\bf w}(\omega) = - \left( \mu + \frac{1}{2} D_0 \omega^2 \right) 
\sigma_3 {\bf w}(\omega) + \int_{-\infty}^{\infty} d\omega_1 
\left( 2 K_1(\omega,\omega_1) \sigma_3 + K_2(\omega,\omega_1) \sigma_2 \right) 
{\bf w}(\omega_1),
\end{equation}
where 
$$
\sigma_3 = \left[ \begin{array}{cc} 1 & 0 \\ 0 & -1 \end{array} \right], 
\;\;\;\;\sigma_2 = \left[ \begin{array}{cc} 0 & 1 \\ -1 & 0 \end{array} 
\right],
$$
and the integral kernels are 
\begin{eqnarray*}
K_1(\omega,\omega_1) & = & \int_{-\infty}^{\infty} d \omega_2 
r[(\omega-\omega_1)(\omega-\omega_2)] \Phi(\omega_2) \Phi(\omega_1+\omega_2-\omega), \\
K_2(\omega,\omega_1) & = & \int_{-\infty}^{\infty} d \omega_2 
r[(\omega_2-\omega_1)(\omega_2-\omega)] \Phi(\omega_2) \Phi(\omega_1-\omega_2+\omega).
\end{eqnarray*}
Using the stationary solution $\Phi(\omega)$ from the previous subsection and 
implementing the Simpson's integration method again, we solve the linear problem 
by using the linear algebra packages built in Matlab 5.2. We identify two types 
of modes of the linear spectrum: symmetric eigenfunctions, when ${\bf w}(-\omega) 
= {\bf w}(\omega)$, and anti-symmetric eigenfunctions, when ${\bf w}(-\omega) = 
- {\bf w}(\omega)$. 

The linear spectrum for the soliton of Fig. 5(a) is shown in 
Fig. 6(a,b). It consists of three main parts: (i) continuous spectrum, 
(ii) neutral (zero) modes, and (iii) internal (oscillatory) modes. 

\begin{figure}
\begin{minipage}{10cm}
\begin{center} \hspace*{-0.5cm} (a) \end{center}
\vspace{-2cm}
\rotatebox{0}{\resizebox{8cm}{10cm}{\includegraphics[0in,0.5in]
 [8in,10.5in]{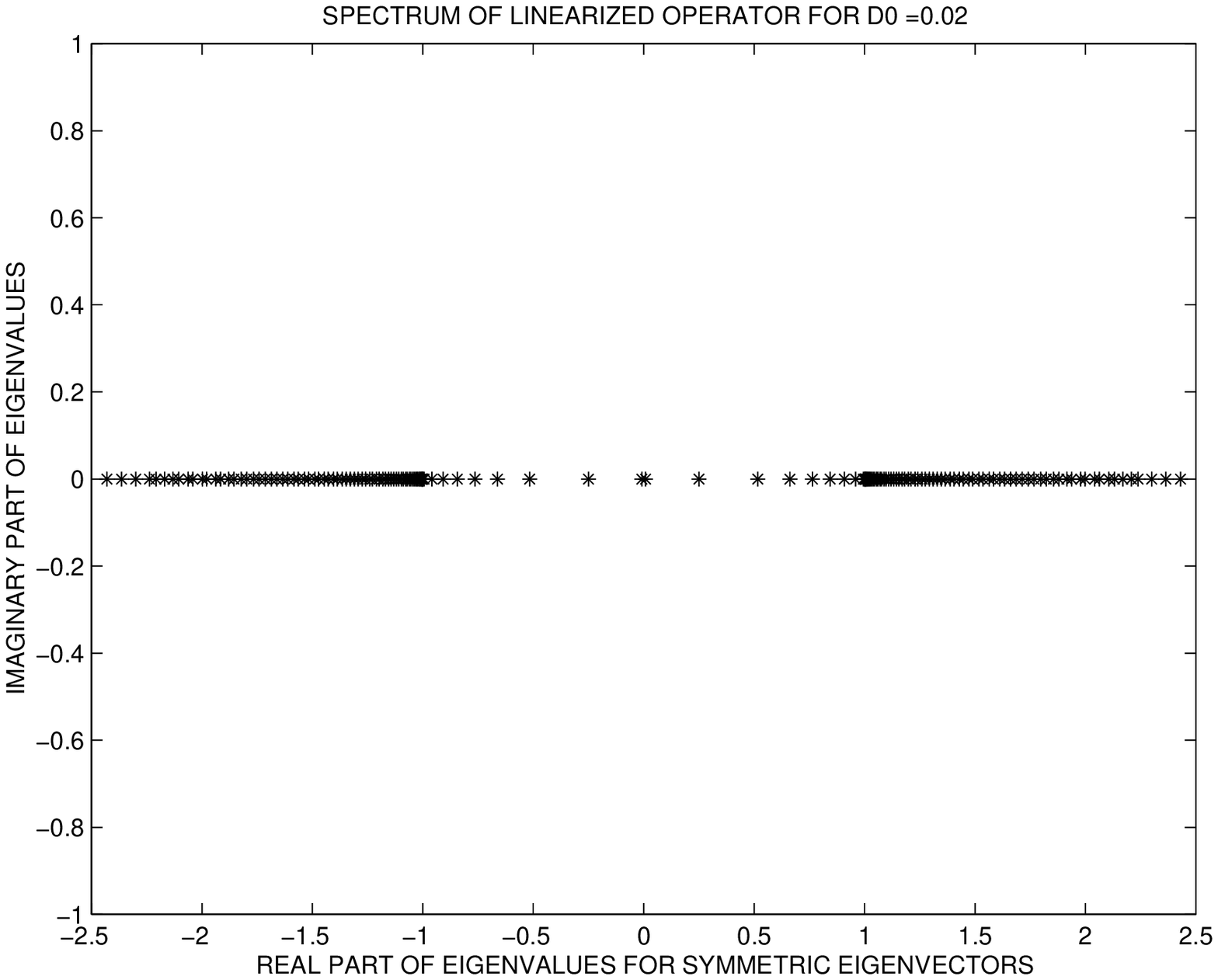}}}
\end{minipage}
\hspace{-1.5cm}
\begin{minipage}{10cm}
\begin{center} \hspace*{-0.5cm} (b) \end{center}
\vspace{-2cm}
\rotatebox{0}{\resizebox{8cm}{10cm}{\includegraphics[0in,0.5in]
 [8in,10.5in]{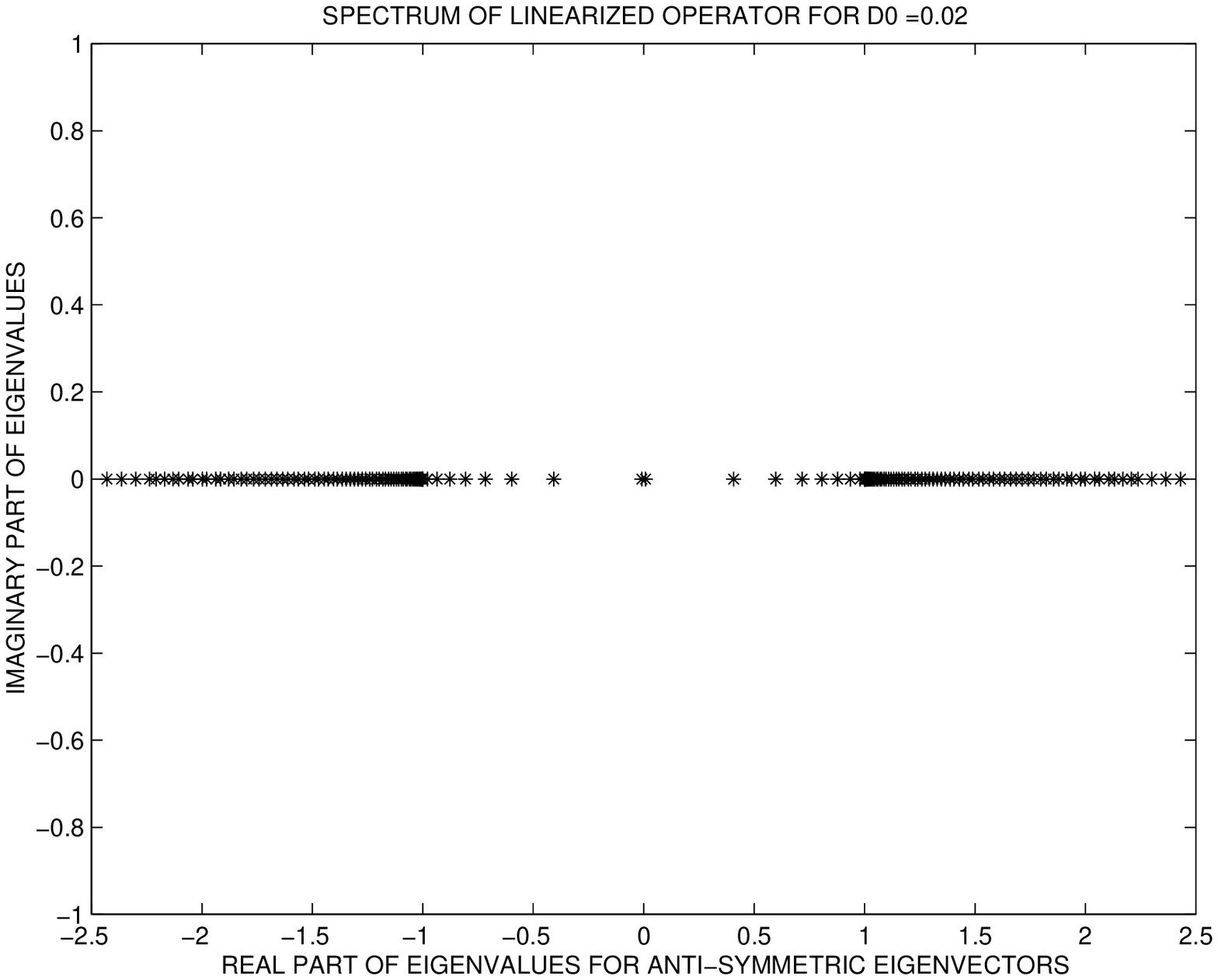}}}
\end{minipage}
\vspace{-1.5cm}
\caption{(a,b) The linear spectrum $\lambda$ of Eq. (\ref{linear_problem}) 
for the stationary pulse at $D_0 = 0.02$, $m = 2$ and $\mu = 1$ which 
corresponds to Fig. 5(a). }
\end{figure}

\begin{figure}
\begin{minipage}{10cm}
\begin{center} \hspace*{-0.5cm} (a) \end{center}
\vspace{-2cm}
\rotatebox{0}{\resizebox{8cm}{10cm}{\includegraphics[0in,0.5in]
 [8in,10.5in]{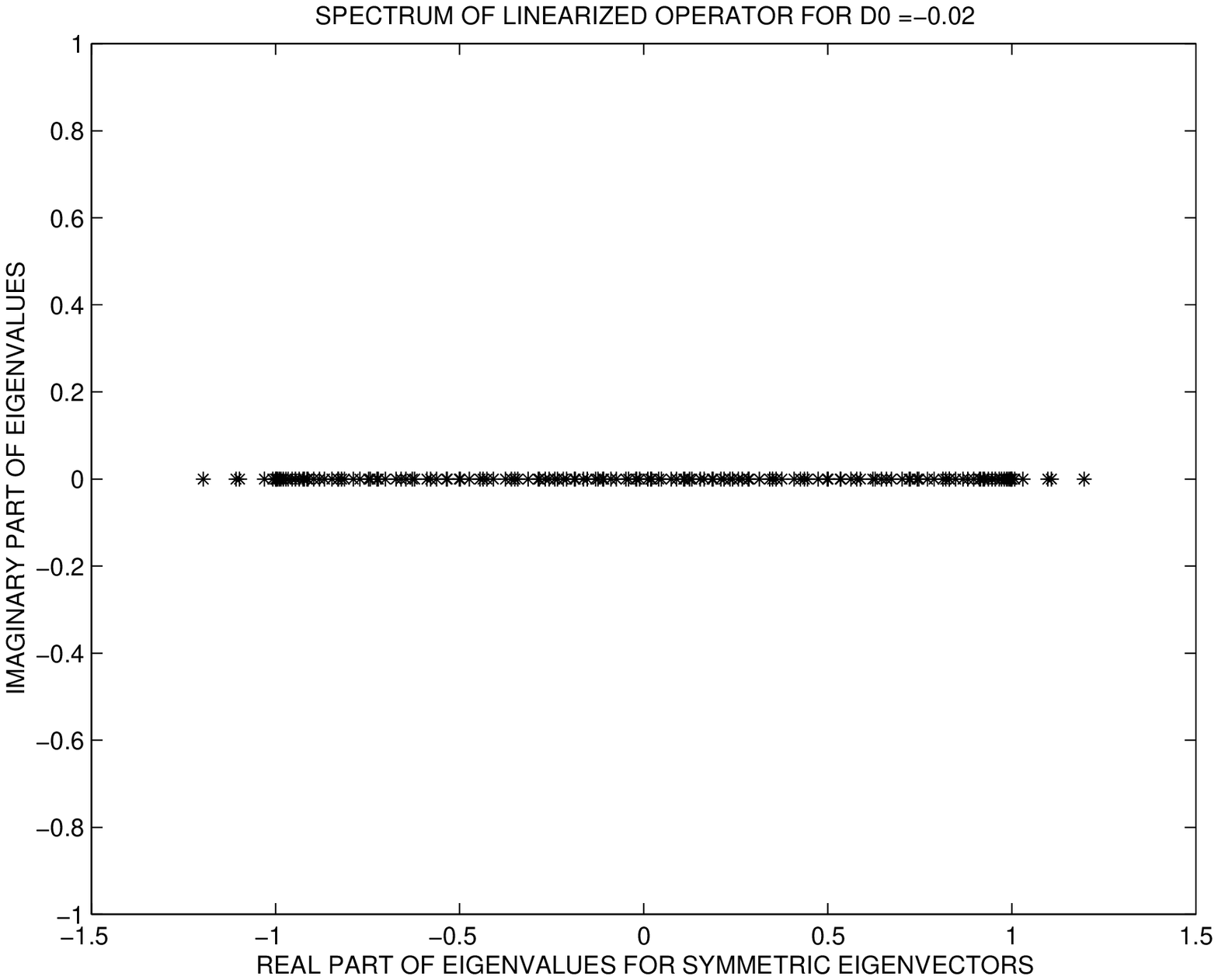}}}
\end{minipage}
\hspace{-1.5cm}
\begin{minipage}{10cm}
\begin{center} \hspace*{-0.5cm} (b) \end{center}
\vspace{-2cm}
\rotatebox{0}{\resizebox{8cm}{10cm}{\includegraphics[0in,0.5in]
 [8in,10.5in]{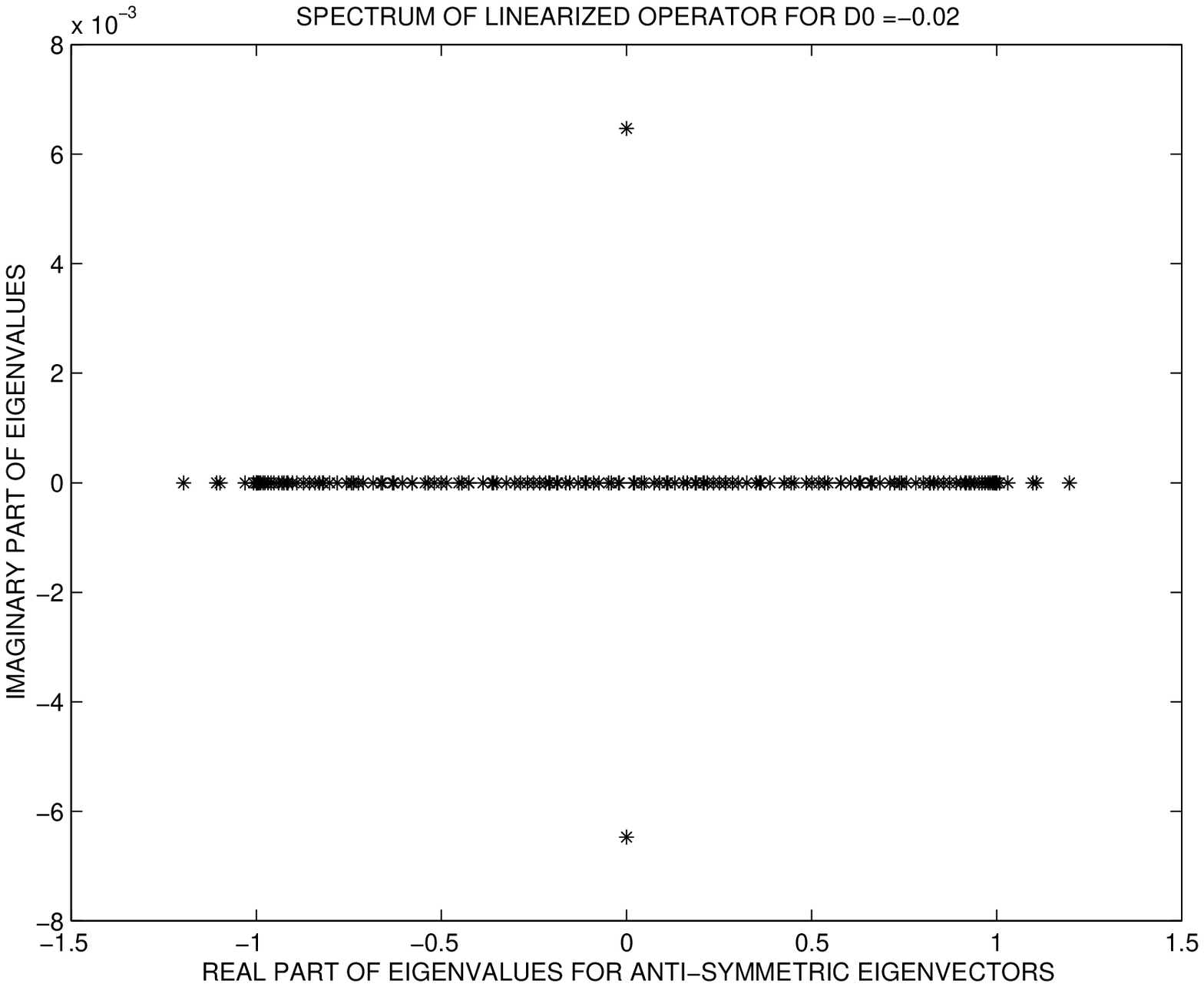}}}
\end{minipage}
\vspace{-1.5cm}
\caption{(a,b) The linear spectrum $\lambda$ of Eq. (\ref{linear_problem}) 
for the stationary pulse at $D_0 = -0.02$, $m = 2$ and $\mu = 1$ which 
corresponds to Fig. 5(b). }
\end{figure}

When $D_0 > 0$ (the anomalous regime of the dispersion map), the continuous 
spectrum is located at the real axis for $|\lambda| > \mu$ 
(see Fig. 6(a,b) where $\mu = 1$). Indeed, the continuous modes are singular in the 
Fourier representation, i.e. ${\bf w}(\omega) \sim \delta(\omega-\Omega)$. Then, 
the linear problem (\ref{linear_problem}) has the continuous spectrum at 
$\lambda = \pm \lambda_{\Omega}$, where 
\begin{equation}
\label{continuous_spectrum}
\lambda_{\Omega} = \mu + \frac{1}{2} D_0 \Omega^2,
\end{equation}
provided the following integral kernels are not 
singular,
\begin{eqnarray*}
\lim_{\omega \to \Omega} K_1(\omega,\Omega-\omega) & = & \frac{1}{4 \pi^2} 
\int_{-\infty}^{\infty} d\omega \Phi^2(\omega), \\
\lim_{\omega \to \Omega} K_2(\omega,\Omega-\omega) & = & \int_{-\infty}^{\infty} 
d\omega r(\omega^2) \Phi(\omega+\Omega) \Phi(\omega-\Omega).
\end{eqnarray*}

The neutral (zero) modes always appear at $\lambda = 0$ as double degenerate 
states for both symmetric and asymmetric eigenfunctions. However, the inaccuracy 
of the numerical method destroys the degeneracy of the zero modes. As a result, 
the two zero modes may appear either for small real or for small imaginary values 
of $\lambda$. Fig. 7(b) displays two imaginary eigenvalues of order of 
${\rm O}(10^{-2})$ which appear to be shifted from the origin of $\lambda$ 
due to this numerical effect. Since the zero modes are not of interest from 
stability point of view, we neglect this effect and leave the scheme without any 
additional modification.

\begin{figure}
\begin{minipage}{10cm}
\begin{center} \hspace*{-0.5cm} (a) \end{center}
\vspace{-2cm}
\rotatebox{0}{\resizebox{8cm}{10cm}{\includegraphics[0in,0.5in]
 [8in,10.5in]{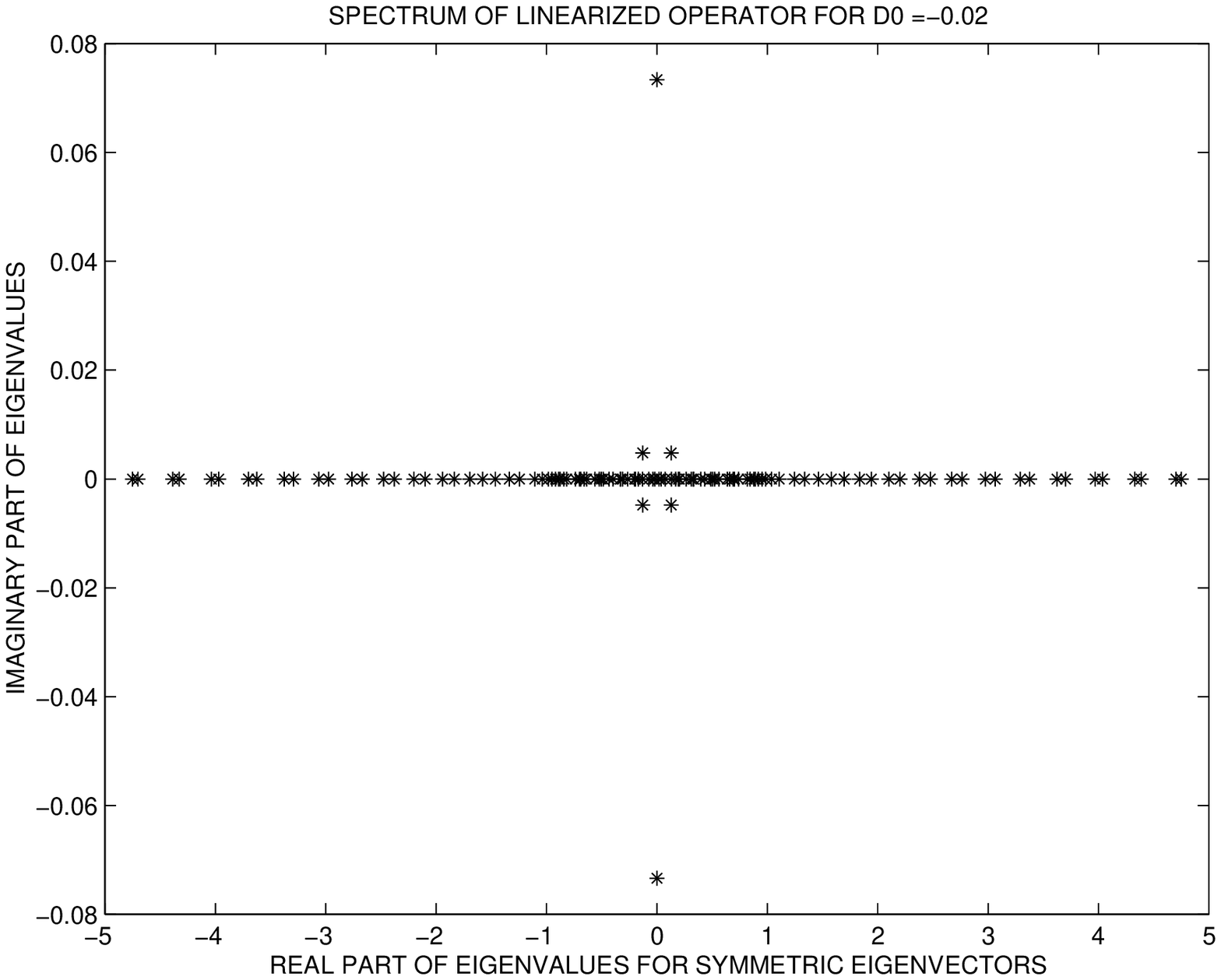}}}
\end{minipage}
\hspace{-1.5cm}
\begin{minipage}{10cm}
\begin{center} \hspace*{-0.5cm} (b) \end{center}
\vspace{-2cm}
\rotatebox{0}{\resizebox{8cm}{10cm}{\includegraphics[0in,0.5in]
 [8in,10.5in]{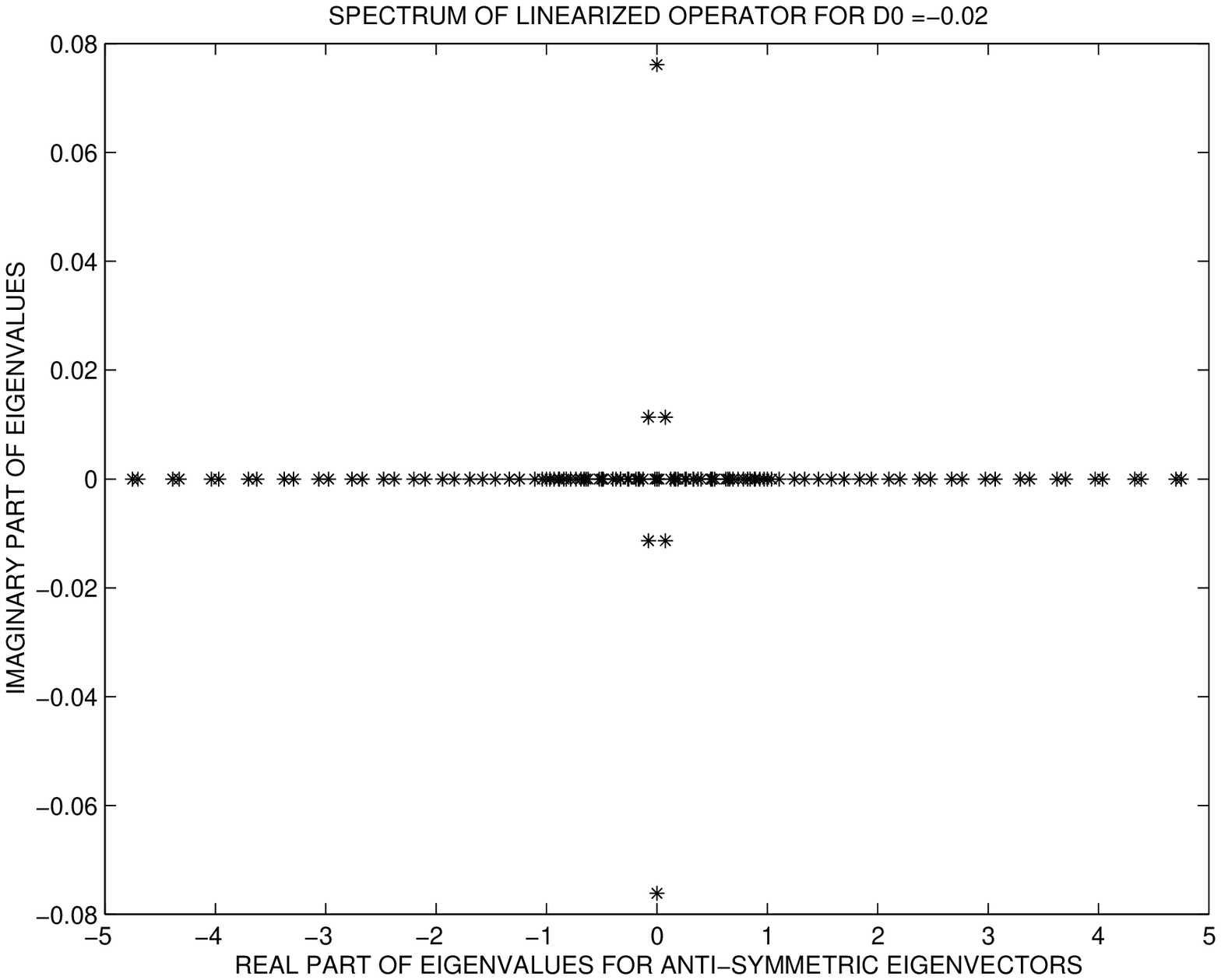}}}
\end{minipage}
\begin{minipage}{10cm}
\begin{center} \hspace*{-0.5cm} (c) \end{center}
\vspace{-2cm}
\rotatebox{0}{\resizebox{8cm}{10cm}{\includegraphics[0in,0.5in]
 [8in,10.5in]{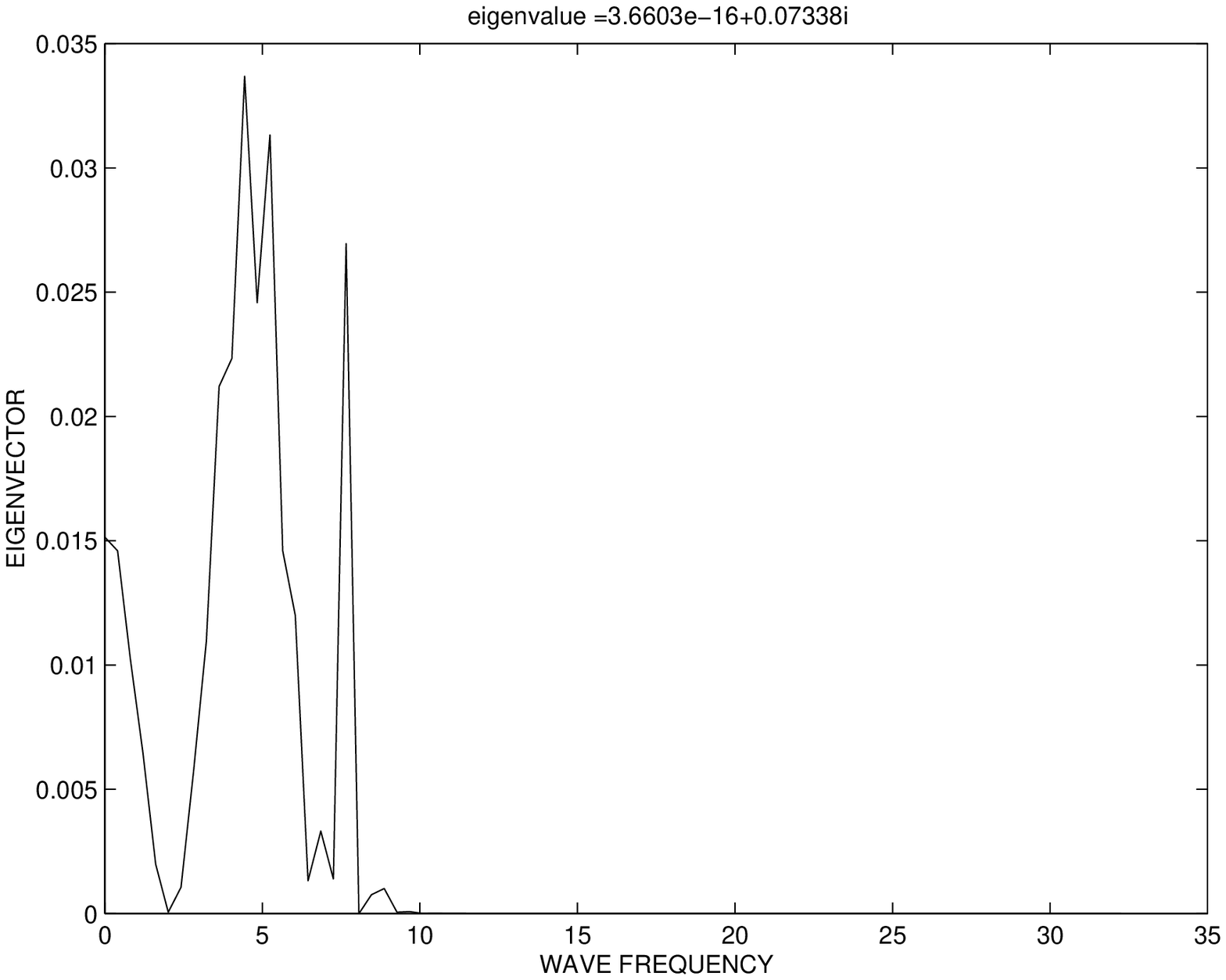}}}
\end{minipage}
\hspace{-1.5cm}
\begin{minipage}{10cm}
\begin{center} \hspace*{-0.5cm} (d) \end{center}
\vspace{-2cm}
\rotatebox{0}{\resizebox{8cm}{10cm}{\includegraphics[0in,0.5in]
 [8in,10.5in]{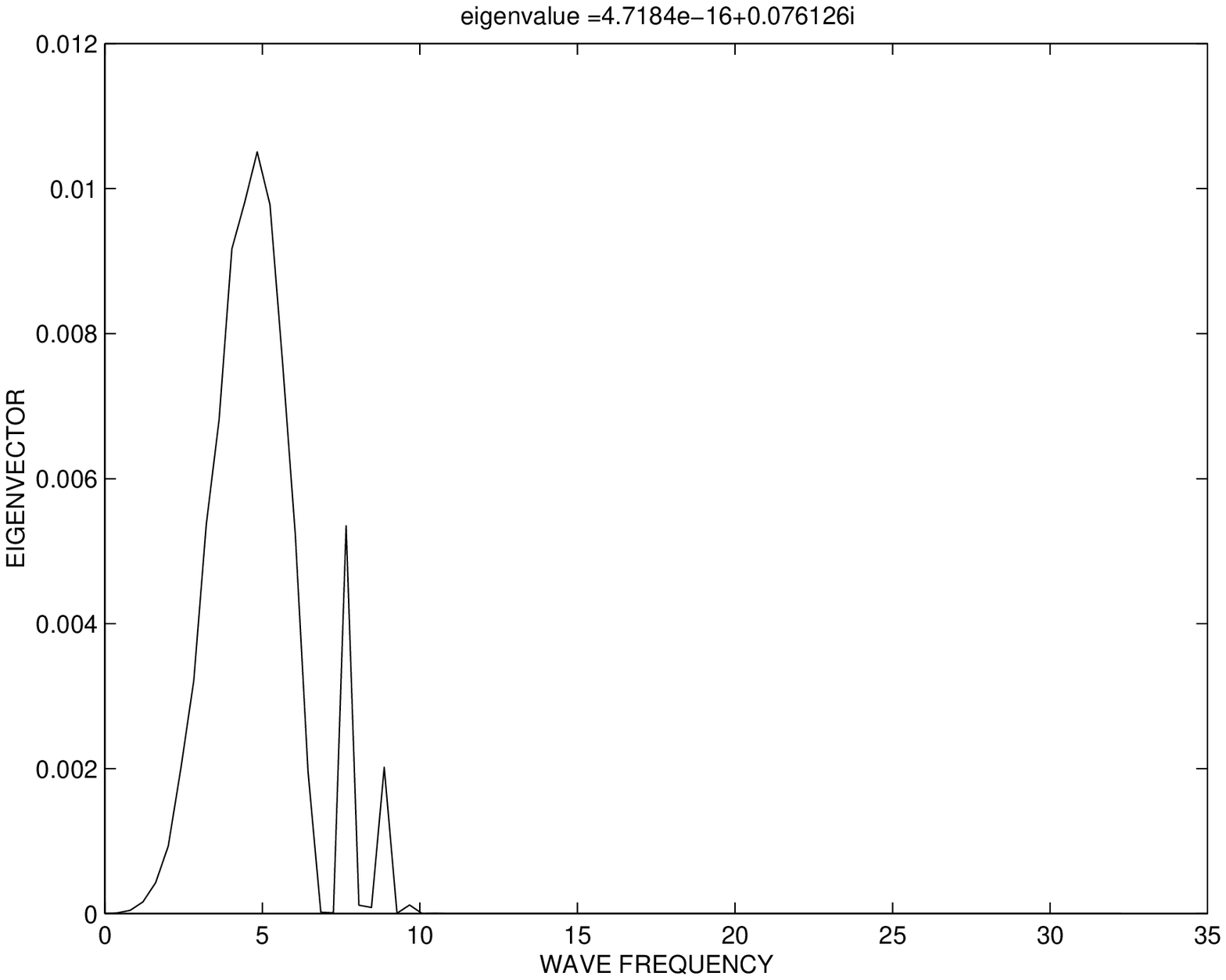}}}
\end{minipage}
\vspace{-1.5cm}
\caption{(a,b) The linear spectrum $\lambda$ of Eq. (\ref{linear_problem}) 
for the stationary pulse at $D_0 = -0.02$, $m = 2$ and $\mu = 1$ which 
corresponds to the Gaussian pulse (\ref{Gaussian}) at branch II. 
(c,d) The symmetric (c) and antisymmetric (d) eigenvectors ${\bf w}(\omega)$ 
for the unstable eigenvalues with ${\rm Im}(\lambda) > 0$. }
\end{figure}

The internal (oscillatory) modes are located in the gap of the continuous spectrum 
as $0 < |\lambda| < \mu$. The set of internal modes may contain different number of 
eigenvalues. We have shown in previous paper \cite{prev} that the set is empty in 
the NLS limit (which corresponds to the limit $\mu \to 0$ at fixed $D_0 > 0$ and $m$). 
[Note that in \cite{prev} we used the Gaussian pulse (\ref{initial}) for 
approximating $\Phi(\omega)$ while here we substitute the numerical result from 
Eq. (\ref{nonlinear}).] Then, we showed that the number of internal modes increases 
with the map's strength (if the parameter $\alpha_s$ is set to unity, the map's 
strength is proportional to $m$ and vice versa). In Fig. 6(a,b) for $\mu = 1$, 
we identify 14 internal modes for symmetric eigenfunctions and $12$ internal modes 
for anti-symmetric eigenfunctions. Still complex eigenvalues are absent for $D_0 > 0$ 
which confirms stability of dispersion-managed solitons in the anomalous regime. 

The linear spectrum for the soliton of Fig. 5(b) is shown in Fig. 7(a,b). 
The continuous spectrum is seen to have changed drastically. When $D_0 < 0$ 
(normal regime of the dispersion map), the continuous spectrum covers the segment 
$| \lambda | < \mu$ twice according to (\ref{continuous_spectrum}). 
As a result, neutral and internal modes, if any, become embedded 
in the wave continuum as seen in Fig. 7(a,b). This indicates a 
resonance of stationary soliton with the linear spectrum in the 
normal regime of the dispersion map. However, this resonance does 
not result to any instability of solitons of branch I within the 
linear theory. We discuss the resonance issue in Section 4. The two 
imaginary eigenvalues on Fig. 7(b) appear from the origin as artifacts 
of the numerical scheme as it was explained above.

At last, we would like to construct the linear spectrum for the solitons of 
branch II. However, the stationary solutions were not identified within the 
Petviashvili's numerical method. Therefore, assuming that the solutions still 
exist, the profile $\Phi(k)$ can only be approximated by the Gaussian pulse 
(\ref{initial}) as we did in \cite{prev}. The linear spectrum in this approximation 
is shown in Fig. 8(a,b) for $D_0 = -0.02$, $m = 2$, and $\mu = 1$. The same type of 
the continuous spectrum is clearly seen not to possess any gap in the origin. 
In addition to this, we identify new complex eigenvalues both for symmetric and 
anti-symmetric eigenfunctions. These complex eigenvalues has relatively large, 
order of ${\rm O}(10^{-1})$, imaginary part and they are associated with the 
instability of the solitons of branch II. The numerical result for the instability 
eigenvalues is in a reasonable comparison with the asymptotic predictions which 
follow from the dynamical model (\ref{alpha0}) and (\ref{beta0}) (see Fig. 3, branch II). 
The eigenvectors ${\bf w}(\omega)$ for the unstable (imaginary) eigenvalues are shown 
in Fig. 8(c,d) for the symmetric and anti-symmetric eigenfunctions, respectively. 
The numerical approximations of the eigenvectors, being inaccurate in details, display 
clearly that the unstable modes are localized at the intermediate wave frequencies 
$\omega$ of the pulse spectrum (at $\omega \approx 4.5$). Thus, the development of the 
unstable eigenvectors would affect the duration of the soliton pulse in the nonlinear 
stage as described in the next subsection. 

\subsection{Non-stationary evolution}

To confirm the transition scenario, we simulate the non-stationary dynamics 
of unstable solitons in the integral model (\ref{integral_model}) by using 
the central-difference scheme,
\begin{equation}
\label{scheme}
\frac{V^{(n+1)}(\omega) - V^{(n-1)}(\omega)}{2 \Delta \zeta} = i 
\int\!\int_{-\infty}^{\infty} d\omega_1 d\omega_2 r(\omega_1 \omega_2) 
e^{i D_0 \omega_1 \omega_2 \zeta_n} V^{(n)}(\omega+\omega_1) 
V^{(n)}(\omega+\omega_2) \bar{V}^{(n)}(\omega+\omega_1+\omega_2) 
\end{equation}
where 
$$
V^{(n)}(\omega) = W(\omega,\zeta_n) \exp\left( \frac{i}{2} D_0 \omega^2 \zeta_n \right)
$$
and $\zeta_n = n \Delta \zeta$. An initial iteration can be done within a forward 
scheme starting with the initial Gaussian pulse (\ref{initial}) with the parameter 
$\alpha_s$ and $E$ corresponding to branches I and II at $\mu = 1$ on Fig. 2(a,b). 
Evolution of a stable long pulse (branch I) is shown in Fig. 9(a,b), while that of 
an unstable short pulse (branch II) is shown in Fig. 9(c,d). The stable long Gaussian 
pulse quickly transits to the stationary pulse given by Eq. (\ref{soliton}) 
(Fig. 5(b)) which propagates later without visible distortions. The second and third 
peeks in the signal spectrum and profile (Fig. 9(a,b)) appear in complete agreement 
with the profile of the stationary DM soliton (cf. Fig. 1 from \cite{AB}). Some oscillations 
along the distance $\zeta$ are excited due to the difference between the Gaussian pulse 
(\ref{initial}) and the exact stationary solution of Eq. (\ref{nonlinear}). These 
oscillations are small compared to the soliton profile and they do not change the 
duration of the soliton pulse (see Fig. 9(a)). The nonlinear resonance at $\omega_{res}$ 
($\omega_{res} = 10$ for Fig. 9(a)) is not seen to be excited during the signal 
propagation. 

\begin{figure}
\begin{minipage}{10cm}
\begin{center} \hspace*{-0.5cm} (a) \end{center}
\vspace{-2cm}
\rotatebox{0}{\resizebox{8cm}{10cm}{\includegraphics[0in,0.5in]
 [8in,10.5in]{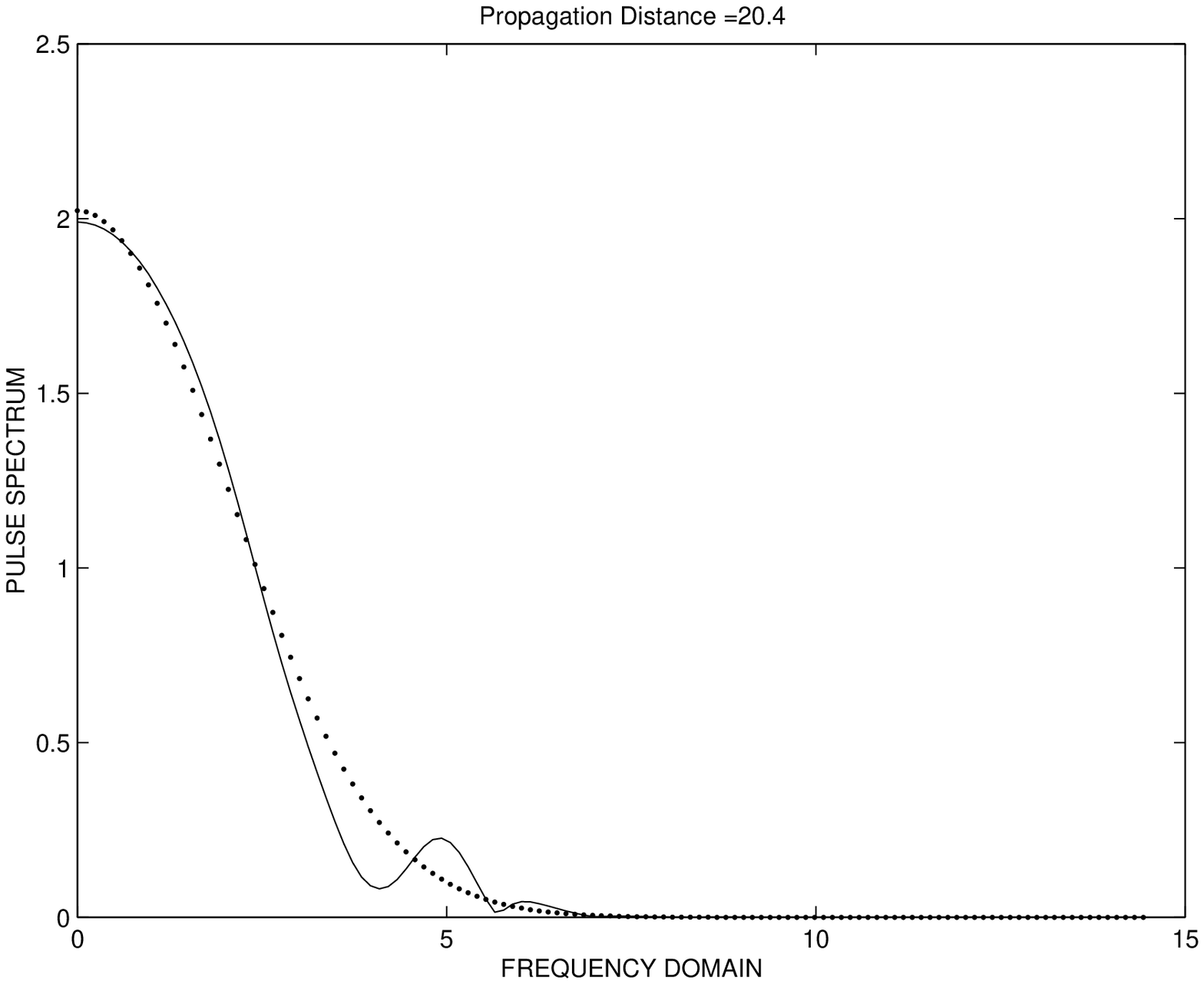}}}
\end{minipage}
\hspace{-1.5cm}
\begin{minipage}{10cm}
\begin{center} \hspace*{-0.5cm} (b) \end{center}
\vspace{-2cm}
\rotatebox{0}{\resizebox{8cm}{10cm}{\includegraphics[0in,0.5in]
 [8in,10.5in]{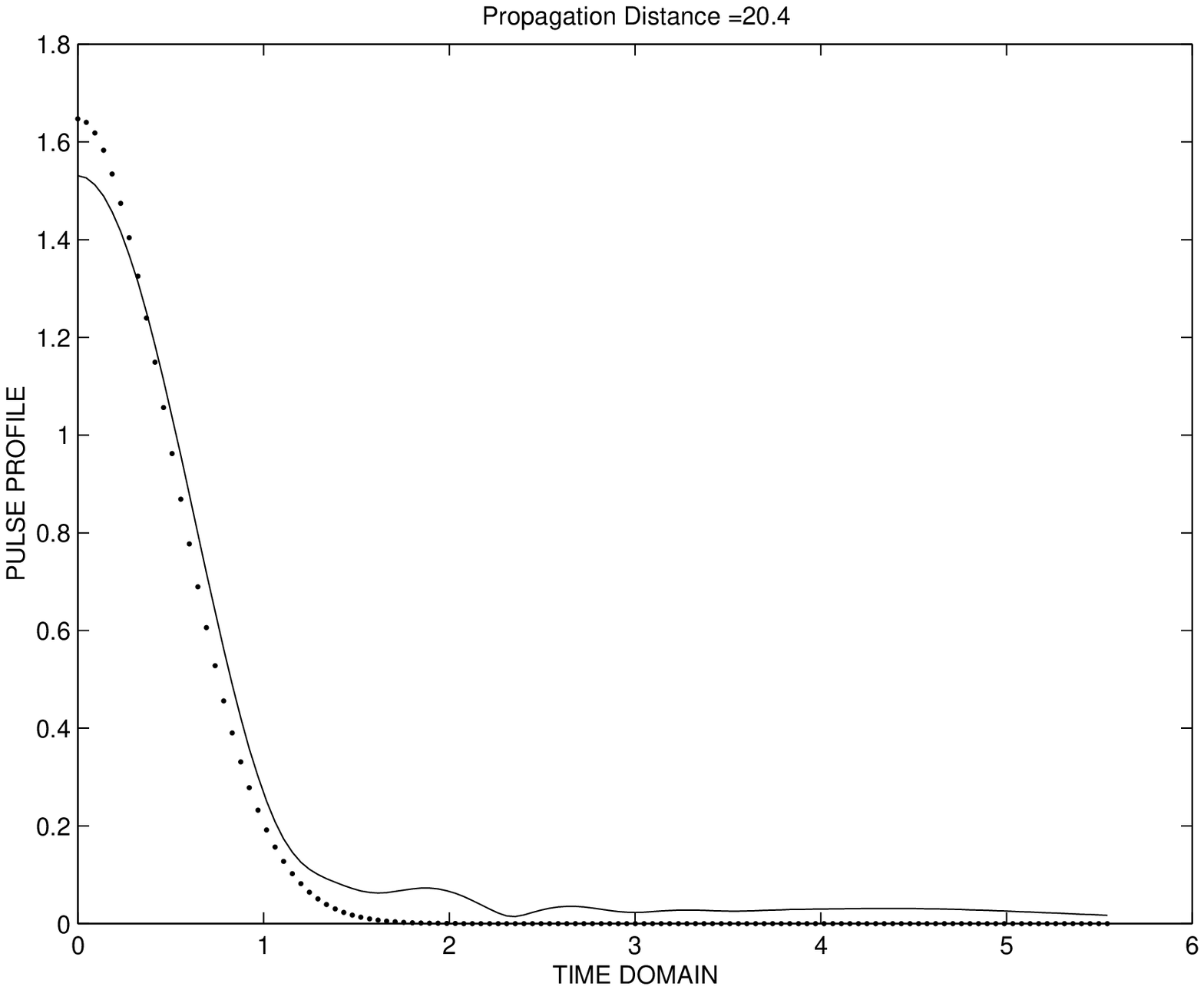}}}
\end{minipage}
\vspace{-1cm}
\begin{minipage}{10cm}
\begin{center} \hspace*{-0.5cm} (c) \end{center}
\vspace{-2cm}
\rotatebox{0}{\resizebox{8cm}{10cm}{\includegraphics[0in,0.5in]
 [8in,10.5in]{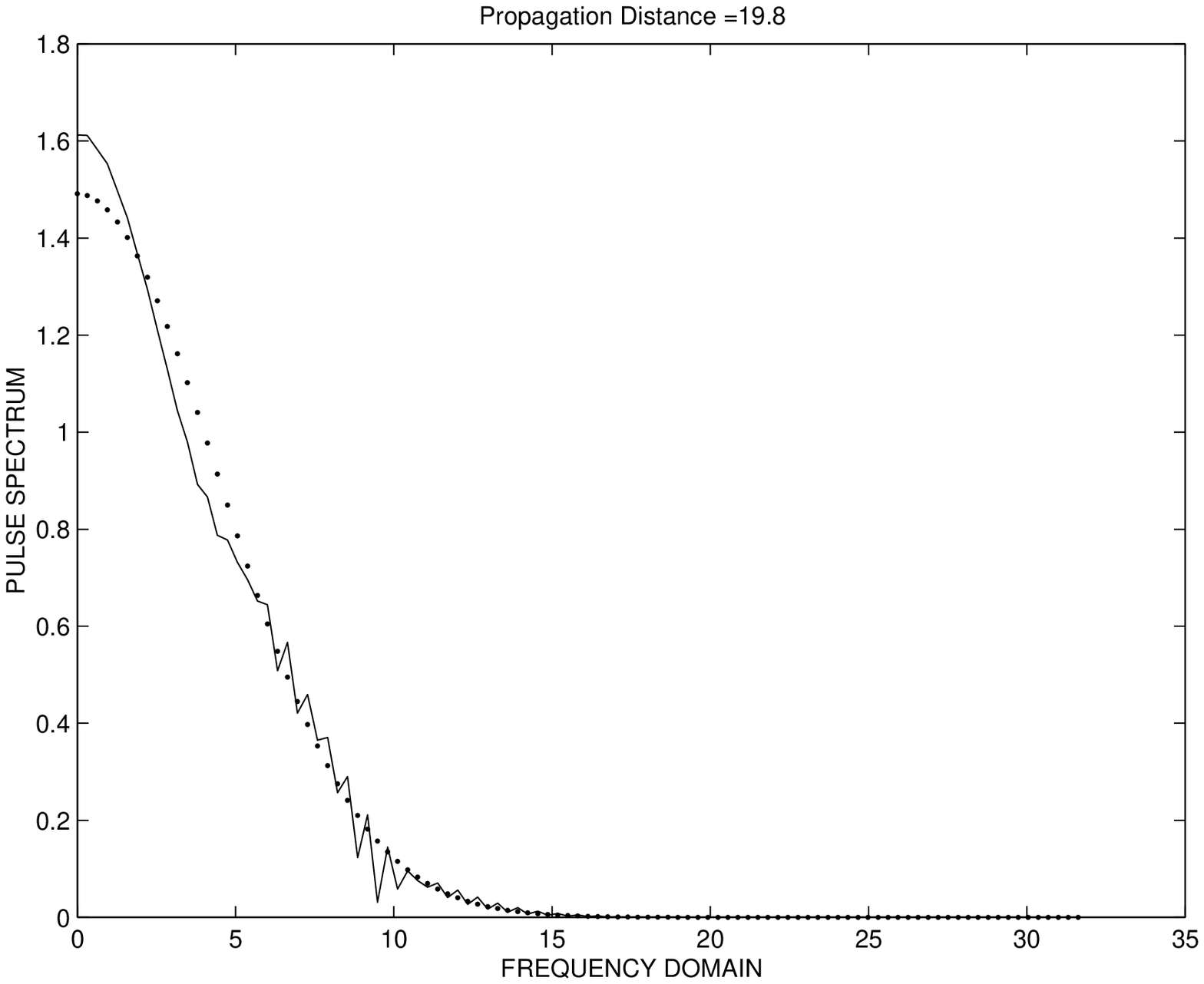}}}
\end{minipage}
\hspace{-1.5cm}
\begin{minipage}{10cm}
\begin{center} \hspace*{-0.5cm} (d) \end{center}
\vspace{-2cm}
\rotatebox{0}{\resizebox{8cm}{10cm}{\includegraphics[0in,0.5in]
 [8in,10.5in]{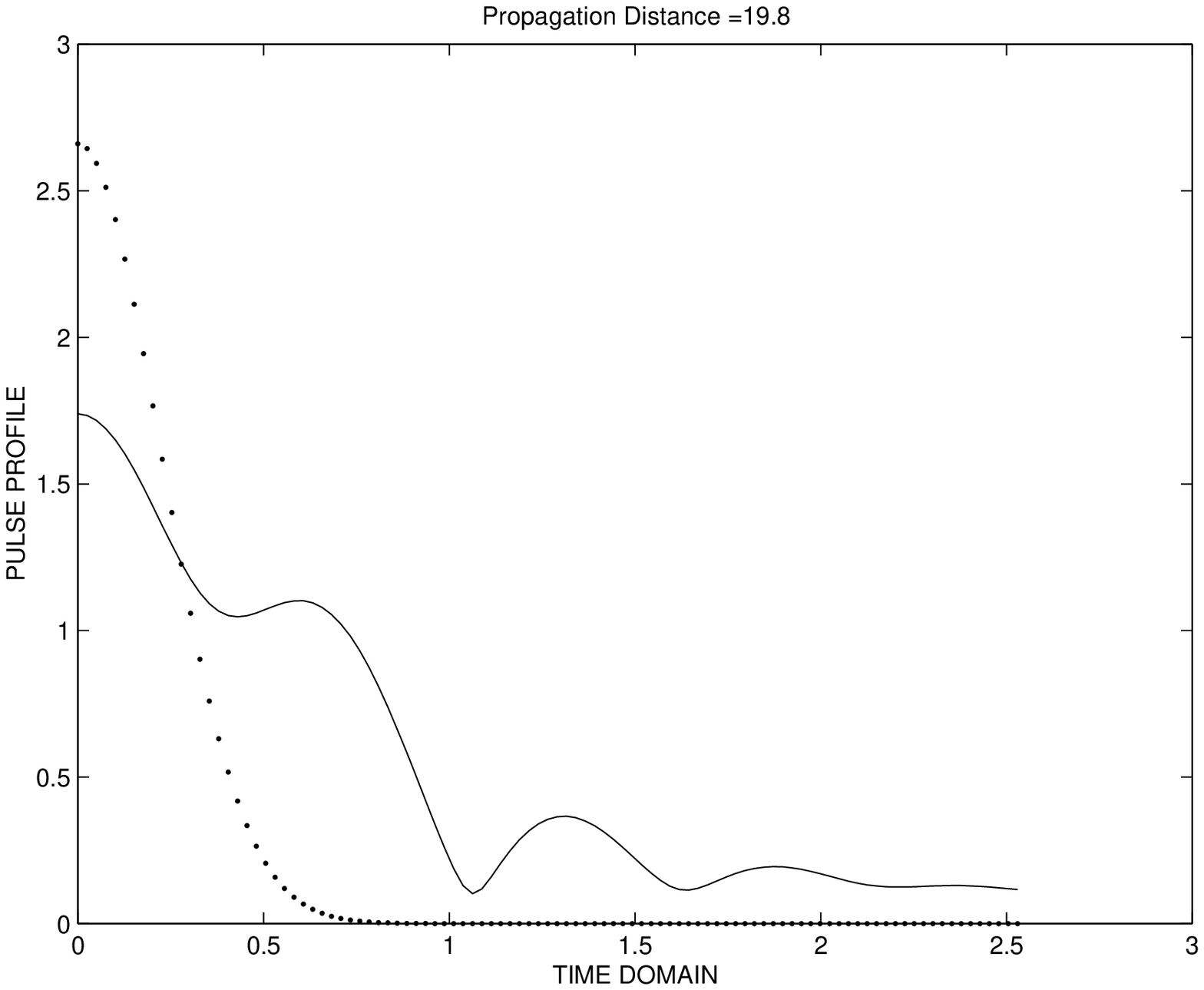}}}
\end{minipage}
\vspace{-1.5cm}
\caption{Propagation of the stable (a,b) and unstable (c,d) Gaussian pulses in the model 
(\ref{integral_model}) at $D_0 = -0.02$ and $m = 2$. The dotted line displays the 
initial pulse (\ref{initial}) for the branch I (a,b) and for the branch II (c,d) at $\mu = 1$. 
The solid line displays the profile at $\zeta = 20$.}
\end{figure}

Evolution of the short (unstable) pulse differs drastically from the previous picture. 
The pulse is being broaden during the evolution, it generates the strong radiation 
tail and tends to the long (stable) signal which has the first node at $\omega 
\approx 4.5$ (cf. Fig. 9(a) and Fig. 9(c)). This transformation is accompanied by 
the intermediate oscillations around the soliton's shape. Thus, we confirm the 
analytical predictions that the short pulses are linearly unstable and switch into 
long stable solitons via long-term oscillatory dynamics. The unstable eigenvectors 
at Fig. 8(c,d) clearly match at Fig. 9(c) with the growing deformations of the 
localization of the pulse spectrum. 

\section{Discussion: Resonance of dispersion-managed solitons}

We have shown that both the Gaussian variational approximation and the integral 
model prescribe the instability of short nonlinear signals in the normal regime of 
the dispersion map. This instability broadens the signal's profile through some 
intermediate oscillations. The long signals propagate stably then. 

We point out that the small-amplitude approximation considered here corresponds to 
the asymptotic limit $E/D_0 \to 0$  (see branch B at Fig. 1 in \cite{GrMen}). 
The stability of the small-amplitude branch for $D_0 < 0$ was previously under 
discussion in the literature \cite{Dor1,GrMen,TurHG}. This discussion can be resolved 
if one takes into account the normalization condition (\ref{normalization}), 
which was used in the previous works. We have checked that the normalization 
condition (\ref{normalization}) selects the pulse solution along the stable branch I 
for $S_{thr} < S < S_{stab}$, where $S_{stab} \approx 6.76$. For $S > S_{stab}$, 
it selects solutions along the unstable branch II (see Figs. 2(a,b)). 
Thus, for the intermediate map strengths (when $S < S_{stab}$), the soliton signal 
propagates stably in the limit of small energies, as reported in \cite{GrMen}. 
However, if the map strength $S$ exceeds the value $S_{stab}$, the soliton signal breaks 
down and switches to a longer signal, as conjectured in \cite{Dor1}. 

For both the branches, we observe the resonance appearing between the linear wave 
spectrum and the stationary signals. This resonance is related to the fact that 
the origin $\lambda = 0$ is absorbed in the continuous spectrum of the linear 
problem. Another way to find the resonance is to construct the linear spectrum 
for the integral model (\ref{integral_model}), $W(\omega,\zeta) \sim W_0 
e^{i \Omega(\omega) \zeta}$, where 
$$
\Omega = - \frac{1}{2} D_0 \omega^2 \geq 0
$$
and $D_0 < 0$. Therefore, for any $\mu > 0$ there exists $\omega_{res}$ such that 
$\Omega(\omega_{res}) = \mu$. This resonance does not show up in the linear theory 
since the discrete and continuous spectra are separated. However, in the nonlinear 
stage, the resonance generally leads to emission of wave packets and soliton's decay. 

Since the transformation or decay of long stable solitons for $D_0 < 0$ have never been 
observed numerically (nor in our simulations reported in Figs. 9), it is likely 
that the effective gap in the spectrum still appears in the nonlinear theory. Also the 
truncated approximation given by the integral model (\ref{integral_model}) may not be 
valid for a correct description of the resonance. The latter issues remain open for 
further analytical consideration.

\section*{Aknowledgements.} 
The author thanks S. Turitsyn and G. Biondini for bringing the present problem to his 
attention as well as T. Lakoba and C. Sulem for collaboration and valuable remarks. 
Fig. 9 was prepared by using computational capacities of Institute of Optics at Rochester, NY.

\end{document}